\documentclass[preprint,3p,11pt,times,numbers,sort&compress]{elsarticle}
\pdfoutput=1

\journal{Journal of Computational Physics}

\usepackage{amsmath}
\usepackage{amsfonts}
\usepackage{amsthm}
\usepackage{hyperref}
\usepackage[version=4]{mhchem}
\usepackage{floatrow}
\usepackage[caption=false]{subfig}
\makeatletter
\providecommand{\doi}[1]{%
  \begingroup
    \let\bibinfo\@secondoftwo
    \urlstyle{rm}%
    \href{http://dx.doi.org/#1}{%
      doi:\discretionary{}{}{}%
      \nolinkurl{#1}%
    }%
  \endgroup
}
\makeatother

\makeatletter
\def\@author#1{\g@addto@macro\elsauthors{\normalsize%
    \def\baselinestretch{1}%
    \upshape\authorsep#1\unskip\textsuperscript{%
      \ifx\@fnmark\@empty\else\unskip\sep\@fnmark\let\sep=,\fi
      \ifx\@corref\@empty\else\unskip\sep\@corref\let\sep=,\fi
      }%
    \def\authorsep{\unskip,\space}%
    \global\let\@fnmark\@empty
    \global\let\@corref\@empty  
    \global\let\sep\@empty}%
    \@eadauthor={#1}
}
\makeatother

\usepackage{glstyle}

\newtheorem{property}{Property}

\newtheorem{remark}{Remark}
\theoremstyle{definition}
\newtheorem{defn}{Definition}[section]

\newcommand{\dpdt}[1]{\partial^-_\tau {#1}}
\newcommand{\intV}{\int_{\Omega}}
\newcommand{\dV}{\diff \Omega}

\begin{document}

\begin{frontmatter}

\title{Transient electrohydrodynamic flow with concentration-dependent fluid properties: modelling and energy-stable numerical schemes}

\author[NBI,SINTEF,UIO]{Gaute Linga\corref{glref}}
\ead{linga@nbi.dk}
\author[NBI]{Asger Bolet}
\author[NBI]{Joachim Mathiesen}
\address[NBI]{
Niels Bohr Institute, 
University of Copenhagen, Blegdamsvej 17, 
DK-2100 Copenhagen, Denmark.}
\address[SINTEF]{SINTEF Digital, Mathematics and Cybernetics, Oslo, Norway}
\address[UIO]{The Njord Centre, PoreLab, University of Oslo, Norway}

\cortext[glref]{Corresponding author.}




\date{\today}

\begin{abstract}
  Transport of electrolytic solutions under influence of electric fields occurs in phenomena ranging from biology to geophysics.
  Here, we present a continuum model for single-phase electrohydrodynamic flow, which can be derived from fundamental thermodynamic principles.
  This results in a generalized Navier--Stokes--Poisson--Nernst--Planck system, where fluid properties such as density and permittivity depend on the ion concentration fields.
  We propose strategies for constructing numerical schemes for this set of equations, where the electrochemical and the hydrodynamic subproblems are decoupled at each time step.
  We provide time discretizations of the model that suffice to satisfy the same energy dissipation law as the continuous model. 
  In particular, we propose both linear and non-linear discretizations of the electrochemical subproblem, along with a projection scheme for the fluid flow.
  The efficiency of the approach is demonstrated by numerical simulations using several of the proposed schemes.
\end{abstract}


\end{frontmatter}


\section{Introduction}
Electrokinetic or electrohydrodynamic flow concerns the coupled transport of charged species and fluid flow in the presence of electric fields \cite{bruus2008}.
Such phenomena have gained increasing attention in recent years due to the rise of the fields of micro- \cite{squires2005} and nanofluidics \cite{schoch2008}.
Important technological applications include biomedical lab-on-a-chip devices \cite{lee2000}, electrophoretic separation of  macromolecules such as DNA and RNA \cite{ghosal2006}, battery and fuel cell technology \cite{nielsen2014,nielsen2015}, desalination of water \cite{nikonenko2014}, and the possibility of harvesting of energy due to salinity gradients (``blue energy'') \cite{siria2017}.
Further, electrokinetic effects can be important within geophysics \cite{hiorth2010,hilner2015}, as fluid flow through charged pores induces a \emph{streaming potential} that counteracts the fluid motion and increases the apparent viscosity \cite{pride1991,fiorentino2016a,fiorentino2016b,bolet2018}.
In fluid-saturated porous rocks, large-scale transport can be mediated by electrochemical gradients \cite{plumper2017}.
Upscaling of the pore-scale electrokinetic description to the macroscale remains an important and challenging research topic \cite{allaire2010,schmuck2011,ray2012,allaire2013,schmuck2015,khoa2017}.


Electrohydrodynamics is usually described by coupling incompressible fluid flow, governed by the Navier--Stokes equations, to solute transport, governed by the Nernst--Planck equations, and electrostatics, governed by a Poisson equation, thereby neglecting magnetic forces.
This results in the nonlinearly coupled Navier--Stokes--Poisson--Nernst--Planck (NSPNP) system of equations \cite{dreyer2013}.
Numerical approaches have often aimed for the steady-state solution to the governing equations \cite{mitscha-baude2017,bolet2018}.
To this end, commercial multi-physics software packages (e.g.\ Comsol) are available, and have long been successfully applied to simulate a variety microfluidic systems.
With regard to the transient development of streaming potential, detailed simulations have often been limited to two-dimensional or axisymmetric geometries such as finite-length symmetric channels \cite{mansouri2005,mansouri2007}.
In studies of electroconvection near perm\-selective membranes \cite{zaltzman2007}, both finite element \cite{pham2012} and (pseudo-) spectral methods \cite{demekhin2011,druzgalski13,druzgalski2016} have proven efficient.
Recently, a spectral method was also applied in a study of the interaction between electrokinetics and turbulent drag \cite{ostilla2017}.
In simulations of electrokinetic flow, the electrolyte solutions are usually assumed to be dilute enough for density, viscosity and permittivity to be independent of the local ion concentrations.
The ion mobilities are usually taken to be proportional to the concentrations.

For the separate \emph{subproblems} comprising the NSPNP problem, there exists many efficient numerical methods.
For the Poisson--Nernst--Planck (PNP) problem, efficient approaches have been demonstrated for semi-conductors \cite{chen2003} and biological ion channels \cite{shen2015}, where e.g.~dispersion and size effects of ions can be included.
For transient simulation of the Navier--Stokes equations, projection methods that date back to Chorin \cite{chorin1967,chorin1968} (see also Guermond, Minev, and Shen \cite{guermond2006}), have imparted speedup compared to solving the monolithic problem, since it effectively decouples the computation of velocity and pressure (although at the cost of some reduced accuracy).
For the full NSPNP problem, however, less is certain, but it seems clear that succesful numerical schemes should aim to decouple, at least, the fluid mechanical subproblem from the electrochemical subproblem, and thus take advantage of the progress made in numerically resolving each of these, although a direct combination does not necessarily yield a successful scheme.

In the field of diffuse-interface (or phase-field) methods for two-phase flow, recent years have seen progress in developing \emph{energy-stable} numerical schemes.
Such schemes are appealing because they share a common property with the physical models in the sense that they, in the absence of external driving forces, unconditionally dissipate energy.
(We give a precise definition of this concept in Sec.\ \ref{sec:numerics}.)
Hence, the schemes can be said to be thermodynamically consistent.
Schemes that do not respect this energy law are prone to numerical errors and instabilities near singularities \cite{fjordholm2011,shen2015}, particularly applicable to flows involving sharp gradients such as both two-phase and electrohydrodynamic flow.
Further, the energy laws permit to establish results on the convergence of numerical schemes.
Schemes that require solving the fully coupled (nonlinear) problem implicitly can relatively easily be constructed to satisfy this property, while a splitting stategy introduces additional difficulty \cite{minjeaud2013,guillen-gonzalez2014}.
Notably, Shen and Yang \cite{shen2015} presented linear, decoupled schemes for phase-field models with density contrast, relying in part on a projection method for the NS equations and a stabilization method for the phase-field equation.

The NSPNP system with two chemical species has been extensively studied by, e.g., Prohl and Schmuck \cite{prohl2009,schmuck2009,prohl2010,schmuck_thesis} who considered also the construction of an energy-stable scheme \cite{prohl2010} with a coupling between the PNP and NS subproblems.
Schemes for multi-ion electrohydrodynamics are also available \cite{bauer2012}.
An energy-stable splitting scheme for a thermodynamically consistent model for two-phase electrohydrodynamics \cite{abels2012} was presented and recently elaborated by \citet{metzger2015,metzger2018}.

\subsection{Contributions of this work}
The objective of this paper is twofold.
One is to obtain a generalized, thermodynamically consistent, model for electrohydrodynamics where the density, viscosity, mobilities, and permittivity depend on the ion concentrations.
The second is to construct \emph{decoupled energy-stable} and \emph{linearized} numerical schemes for this model.
To this end, we will consider a general, thermodynamically consistent model for single-phase flow including electric fields and transport of ions, i.e.~a generalized NSPNP system.
The subproblems of fluid flow and electrochemistry will be decoupled, where the key to energy-stability lies in a forward-projected velocity that enters in the advection term in the solute transport equation, an idea which builds heavily on approaches used in two-phase flow models \cite{minjeaud2013,guillen-gonzalez2014,shen2015,metzger2015}.
For the electrochemical suproblem we propose discretization strategies that suffice to satisfy energy stability \cite{prohl2010}, one of which consititutes a linear scheme.
For the fluid-mechanical part we consider two linear approaches, both a coupled strategy and a projection scheme for this subproblem.
To the authors' knowledge, it is the first time an energy-stable projection scheme has been presented for electrohydrodynamic flow, in particular with concentration-dependent densities, viscosities and permittivities.
Our schemes are shown to be numerically convergent by means of an electrohydrodynamic Taylor--Green vortex; to be numerically energy stable by a stress test of ions flowing in a closed container; a reaction cell to test the reliability of the reaction kinetics; and lastly applied to a geophysical setting, a porous media flow, to demonstrate the potential of the schemes in practical simulations.


\subsection{Outline}
The outline of the paper is as follows.
In Sec.\ \ref{sec:model}, we present a derivation of the model for electrohydrodynamic flow that we consider, and in Sec.\ \ref{sec:properties}, we investigate some properties of the resulting model.
In Sec.\ \ref{sec:numerics} we present discretization strategies for the model, i.e.~numerical schemes for the electrochemical and hydrodynamical subproblems.
Further, in Sec.\ \ref{sec:results} we present numerical simulations using combinations of the numerical schemes presented, for the case of the conventional NSPNP model, and in Sec.\ \ref{sec:discussion} we conclude and provide a brief discussion.

\subsection{Notation}
It is useful to present some remarks on notation before we embark on the main part of the paper.
We will denote an integral of a general quantity $f$ over the domain $\Omega$ by
\begin{equation}
  \intV f \, \dV .
\end{equation}
The $L^2$ inner product of the quantities $a$ and $b$ is denoted by $(a, b)$.
For scalars $f$ and $g$, the inner product is defined by
\begin{equation}
  (f, g) = \intV f g \, \dV,
\end{equation}
while for two $d$-dimensional vectors $\v f$ and $\v g$, and two $d \crossproduct d$ matrices $\v F$ and $\v G$, respectively, it is defined by
\begin{equation}
  (\v f, \v g) = \intV \v f \cdot \v g \, \dV \qquad \textrm{and} \qquad
  (\v F, \v G) = \intV \v F : \v G \, \dV .
\end{equation}
Herein, $\v F : \v G = \sum_{i=1}^d \sum_{j=1}^d F_{ij} G_{ij} $ where $F_{ij}$ and $G_{ij}$ denote the components of $\v F$ and $\v G$, respectively.
The $L^2$ norm of a general quantity $a$ is denoted by $\norm{a}$.
In particular,
\begin{equation}
  \norm{f}^2 = (f, f) = \intV |f|^2 \, \dV. 
\end{equation}
The $L^2$ norm of a quantity $a$ over the boundary $\partial\Omega$ is denoted by $\norm{a}_{\partial\Omega}$, that is,
\begin{equation}
  \norm{f}^2_{\partial\Omega} = \int_{\partial\Omega} | f |^2 \, \diff\Gamma
\end{equation}
A general time-discretized quantity $a$ evaluated at the time step $k$ is denoted by a superscript, $a^k$.
For the time discretization strategies in the forthcoming, we will make use of the backwards-differencing discrete differential operator.
For the sake of simplicity, we adopt the following notation for a discrete time derivative:
\begin{equation}
  \dpdt f^k = \frac{f^k-f^{k-1}}{\tau},
  \label{eq:def_dpdt}
\end{equation}
where $f$ is a general function (scalar or vector), and $\tau$ is a discrete time step.

\section{A general model for single-phase electrohydrodynamics}
\label{sec:model}
Physically, single-phase electrohydrodynamic flow consists of the coupled system of fluid flow, ion transport and electrostatics.
Such a continuum modelling approach is realistic down to the scale of a few nanometers.
We will in the coming sections present a derivation, using variational principles, of a thermodynamically consistent and frame-invariant model of electrohydrodynamic flow, where the fluid properties are allowed to depend on the local concentrations of the chemical species.
The main approximation underlying the model is that the volume of a fluid element does not change with increasing concentrations, only the mass, and hence the velocity field can be taken to be solenoidal.
We will end up with the following partial differential equations, evolving in the spatial coordinate $\v x \in \Omega \subset \mathbb{R}^d$, where $\Omega$ is the domain and $d$ is the dimension, and in time $t$:
\begin{align}
  \rho (\{ c_j\}) \pdt {\v u } + (\v m \cdot \grad ) \v u - \div ( 2 \mu (\{ c_j \}) \v D \v u ) + \grad p &= - \sum_i c_i \grad g_i
                                                                                              , \label{eq:model_NS1}\\
  \div \v u &= 0, \label{eq:model_NS2}\\
  \pdt c_i + \v u \cdot \grad c_i &= \div \left( K_i (\{ c_j \}) \grad g_i \right) + R_i
                                    \label{eq:model_ci}\\
                                    g_i &= \sum_j \pd {M_j} {c_i} + z_i V - \pd {\rho}{c_i} \v a_g \cdot \v x - \frac{1}{2} |\grad V |^2 \pd {\epsilon}{c_i},
                                    \label{eq:model_gi}\\
  \div ( \epsilon (\{ c_j \}) \grad V) &= - \sum_i z_i c_i .\label{eq:model_V}
\end{align}
Here, the following quantites are involved.
\begin{enumerate}
\item [$\rho$] --- fluid density,
\item [$\v u$] --- velocity field,
\item [$\v m$] --- advecting momentum (defined below in \cref{eq:ad_mom}),
\item [$\mu$] --- dynamic viscosity,
\item [$p$] --- pressure,
\item [$c_i $] --- concentration of ion species $i \in 1, \ldots, N$,
\item [$g_i$] --- the chemical potential associated with species $i$, 
\item [$K_i$] --- the mobility of species $i$,
\item [$R_i$] --- reaction source term for species $i$,
\item [$M_i$] --- a specific energy related to having ion species $i$ dissolved,
\item [$z_i$] --- valency of species $i$,
\item [$\v a_g$] --- the gravitational acceleration,
\item [$V$] ---  electric potential,
\item [$\epsilon$] --- electric permittivity.
\end{enumerate}
In this general formulation, \emph{the fluid properties $\rho$, $\mu$, $K_i$, $M_i$, and $\epsilon$ are allowed to depend on the set of concentrations $\{c_j\}_{j=1}^N$}, abbreviated as $\{ c_j \}$.
In particular, we assume that the following linear equation of state holds for the density:
\begin{equation}
  \rho(\{ c_j \}) = \rho_0 + \sum_{j=1}^N \pd {\rho}{c_j} c_j .
  \label{eq:eos_rho_cj}
\end{equation}
Here, $\rho_0$ is the density of the ``background'' fluid, typically water, and the constant $\pdinl {\rho} {c_i} = \mathcal M w_i$, where $\mathcal M$ is a constant conversion factor and $w_i$ is the number of nuclei in a given species $j$.
Together with \cref{eq:model_NS2}, \cref{eq:eos_rho_cj} implies that the chemical species that enters into a volume element does not change its volume, but only its local mass.
This is a simplifying assumption, and is distinct from the related model by \citet{dreyer2013} who assume a constant number density to close the equations.

Note that in our formulation of the model \eqref{eq:model_NS1}--\eqref{eq:model_V}, we have reduced the number of parameters to a minimum, such that some prefactors have been absorbed into the relevant variables.

Eqs.\ \eqref{eq:model_NS1} and \eqref{eq:model_NS2} are the Navier--Stokes equations with variable density.
Here, the \emph{advecting momentum}
\begin{equation}
  \v m = \rho \v u - \sum_i \pd {\rho} {c_i} K_i \grad g_i,
  \label{eq:ad_mom}
\end{equation}
differs from the canonical momentum $\rho \v u$ due to mass diffusion and migration through $c_i$.
This reflects the difference between the \emph{mass-averaged} velocity $\v u_{\rm mass} = \v m / \rho$ and the conventional \emph{volume-averaged} velocity $\v u$.
This difference has been noted previously in the literature, as in the series of papers by Brenner \cite{brenner2005navier,brenner2005kinematics,brenner2006fluid}, and is standard for diffuse-interface models for two-phase flow \cite{abels2012,campillo-funollet2012}.

An unconventional forcing term on the right hand side of \eqref{eq:model_NS1}, $- \sum_i c_i \grad g_i$ can by a redefinition of the pressure, and integration by parts, be written as the more conventional
\begin{equation}
  \rho \v a_g - \rho_e \grad V - \frac{1}{2} |\grad V|^2 \grad \epsilon,
\end{equation}
which reveals the origin of the (conservative) driving forces in that may be present in the system.
The terms represent, respectively, gravity, electric force, and a Helmholtz force due to permittivity gradients.
However, the formulation of the right hand side in \eqref{eq:model_NS1} has e.g., numerical advantages, as $g_i$ is constant at equilibrium, and therefore near equilibrium, the term $-\sum_i c_i \grad g_i$ will be less prone to catastrophic cancellation and pressure-buildup in the electric double layer \cite{nielsen2014}.

Further, the symmetric gradient entering into the viscous term is defined by $\v D \v u = \sym(\grad \v u) = (\grad \v u + \grad \v u^\top)/2$.
Eqs.\ \eqref{eq:model_ci} and \eqref{eq:model_gi} can be seen as a generalized Nernst--Planck equation for transport of species.
Finally, \cref{eq:model_V} is the Poisson equation, or Gauss' law for electrostatics, with non-constant permittivity.
Note that \cref{eq:model_V} is expected to be valid provided that charges move sufficiently slow for magnetic forces to be neglected (and thus there is no magnetic body force contribution in \cref{eq:model_NS1}).

\subsection{Typical modelling assumptions}
Typically in electrohydrodynamics, the standard Nernst--Planck equation is used, and the mobility that enters here is then given by $K_i = D_i c_i$, where $D_i$ is the diffusion constant of species $i$.
Herein, the entropic contribution to the specific free energy $M_j$ is classically given by
\begin{equation}
  M_j = c_j ( \ln c_j - 1 ) + \beta_{j} c_j,
  \label{eq:typical_Mj}
\end{equation}
where $\beta_j$ is a constant that can be related to reactions with other species, and shall be elaborated on later, specifically towards the end of Sec.\ \ref{sec:modelling_choices} and in \ref{sec:modelling_reaction_terms}.
Note that in general, as it does in Eq.\ \eqref{eq:typical_Mj}, $M_j$ contains logarithmically divergent terms that require special care both numerically and within stability proofs.
Regularisation of the divergent behaviour in \cref{eq:typical_Mj} is discussed in Sec.\ \ref{sec:chemicalenergyfunction} and in Ref.\ \cite{metzger2018}.

In equilibrium and for constant dielectric permittivity $\epsilon$, Eq.\ \eqref{eq:model_V} together with \eqref{eq:model_ci} and \eqref{eq:model_gi} for two symmetric species of opposing charge ($z_+ = - z_- = z$), will reduce to the standard Poisson--Boltzmann description of electrokinetic phenomena.

\subsection{Boundary conditions}
To close the system, we need to assign boundary conditions on the boundary $\partial\Omega$ of $\Omega$.
\begin{itemize}
\item \textbf{Velocity boundary condition:} On the velocity field $\v u$, we assign a general Navier slip condition (see e.g.\ \cite{aland2016,qian2006}),
\begin{subequations}
  \begin{align}
    \v u \cdot \hat{\v n} &= 0, \label{eq:unBC} \\
    \left[ \lambda \v u + 2\mu \v D \v u \cdot \hat{\v n}\right]  \cdot \hat{\v t} &= \v 0, \label{eq:utBC}
  \end{align}
  \label{eq:NavierSlipBC}
\end{subequations}
which is quantitatively correct at the smallest scales and particularly relevant for superhydrophobic surfaces \cite{barrat1999}.
  Herein, $\lambda(\{c_i\})$ describes the contact line friction, $\hat{\v n}$ is a unit vector pointing out of the domain, and $\hat{\v t}$ is a unit tangent vector.
  The function $\lambda$ is related to the slip length $\ell_{\textrm{slip}}$ through the relation $\ell_{\textrm{slip}} = \mu / \lambda$.
  In the case where $\lambda \to \infty$ ($\ell_{\textrm{slip}} \to 0$), we retrieve the commonly used no-slip condition: 
  \begin{equation}
    \v u = 0, \label{eq:noslip}
  \end{equation}
  which is realistic even on small scales, and extensively used for example in studies of electrokinetic instability \cite{druzgalski13,druzgalski2016} and microchannel flows \cite{nielsen2014}.
  Note that in the numerical simulations presented in Sec.\ \ref{sec:results}, we shall assume that \eqref{eq:noslip} holds.
\item \textbf{Electrostatic potential boundary condition:} On the electrostatic potential $V$, the two condtions
  \begin{equation}
    \hat{\v n} \cdot \epsilon \grad V = \sigma_e, \quad \textrm{and} \quad V = V_0, \label{eq:VBC}
  \end{equation}
  can be applied to separate parts of the domain.
  In Eq.\ \eqref{eq:VBC}, $\sigma_e$ is the assigned surface charge of the boundary, and $V_0$ is a given potential at the boundary.
  While $\sigma_e$ may vary in space, $V_0$ will in the forthcoming be assumed to take a constant value to avoid injecting energy into the closed system (for which we will prove energy stability).
  Without loss of generality, we will thus fix $V_0 = 0$ henceforth.
\item \textbf{Chemical species boundary conditon:} For each chemical species $c_i$, we assume that the diffusive flux across the boundary vanishes, i.e.\
  \begin{equation}
    \hat{\v n} \cdot K_i \grad g_i = 0. \label{eq:gBC}
  \end{equation}
  Together with Eq.\ \eqref{eq:unBC}, Eq.\ \eqref{eq:gBC} represents impenetrable boundaries for the chemical species.
\end{itemize}

\subsection{Reaction terms}
With regard to modelling the reaction terms $R_i$, we consider a sequence of reactions $m \in 1, \ldots, M$, where each reaction $m$ can be written in the compact form
\begin{equation}
  0 \rightleftharpoons \sum_m \nu_{m,i} \chi_i,
\end{equation}
where $\nu_{m,i}$ is the \emph{net} stoichiometric coefficent (products minus reactants) of ion $i$ in reaction $m$, and $\chi_i$ is the chemical symbol of ion $i$.
In \ref{sec:modelling_reaction_terms}, we argue that we can model
\begin{equation}
  R_i = \sum_m \nu_{m,i} \mathcal R_m
  \quad \textrm{with} \quad
  \mathcal{R}_m =- \mathcal{C}_m \cdot \sum_j \nu_{m,j} g_j,
  \label{eq:reaction_R_text}
\end{equation}
where $\mathcal{C}_m \geq 0$ is a function of the involved variables. 
Such modelling of the reaction term was also considered by, e.g., Refs.\ \cite{campillo-funollet2012,metzger2015,metzger2018}.
Note that $\mathcal{C}_m$ can also be a function of the spatial coordinate $\v x$, i.e., a reaction can be promoted or demoted in a certain region of the domain; effectively allowing to simulate, e.g., catalytic or other electrochemical systems.

\subsection{Derivation of the model}
\label{sec:derivation_model}
We now present a derivation of a model for general electrohydrodynamic flow.
The forthcoming analysis is similar to that considered by previous authors \cite{abels2012,campillo-funollet2012}.
We seek to formulate a model where the fluid properties are allowed to depend on the concentrations, which is both \emph{frame-invariant} (Galilei invariant), thermodynamically consistent (dissipates free energy), and where the velocity field is solenoidal (divergence-free).
The latter point limits the generality of the model, in the sense that we consider \emph{quasi-incompressible} fluids; such that the local concentration fields only makes a fluid element heavier, but does not make it expand.
This is a fair assumption for e.g.\ dissolving table salt in water under certain conditions.
In general, however, liquids can both contract and expand with the addition of another component.
Moreover, this behaviour can be non-monotonous.

The evolution of the concentration fields $c_i$ can in general be written as the advection--diffusion--reaction equation
\begin{equation}
  \pdt c_i + \div ( c_i \v u ) = - \div \v J_{c_i} + R_i,
  \label{eq:pd_c_i}
\end{equation}
where $\v J_{c_i}$ is a hitherto undetermined diffusive flux, and $R_i$ is a reaction source term.
The left hand side is for convenience written in the convective form.

For the density field we assume the linear equation of state \eqref{eq:eos_rho_cj}.
With the quasi-incompressible assumption, the velocity field will still, as without any solutes, be solenoidal, i.e., 
\begin{equation}
  \div \v u = 0.
  \label{eq:solenoidal}
\end{equation}
Using \eqref{eq:eos_rho_cj}, \eqref{eq:solenoidal} and \eqref{eq:pd_c_i} we can derive the evolution of the density,
\begin{align}
  \pdt \rho + \div ( \rho \v u ) 
  &= \sum_i \pd {\rho} {c_i} \left[ \pdt c_i + \div ( c_i \v u ) \right] 
  \\
  &= \sum_i \pd {\rho} {c_i} \left[ - \div \v J_{c_i} + R_i \right] 
  \\
  &= - \div \left( \sum_i \pd {\rho} {c_i} \v J_{c_i} 
    \right) 
    ,
\end{align}
or
\begin{equation}
  \pdt \rho + \div ( \rho \v u ) = - \div \v J_{\rho},
  \label{eq:pd_rho}
\end{equation}
where we have used the condition that a reaction does not change the density, i.e., $\sum_i R_i \pdinl {\rho} {c_i}  = 0$.
This follows from the quasi-incompressible condition, and the fact that mass is conserved in a reaction (for all practical purposes, as the binding energy is, as far as these conservation laws are concerned, negligible compared to the rest energy of an atom or molecule).
We have also implicitly defined the diffusive density flux,
\begin{equation}
  \v J_{\rho} = \sum_i \pd {\rho} {c_i} \v J_{c_i} . 
\end{equation}
Eq.\ \eqref{eq:pd_rho} suggests that the mass is transported by the velocity
\begin{equation}
  \v u_{\rm mass} = \v u + \rho^{-1} \v J_{\rho} .
\end{equation}
Following the discussion in Refs.~\cite{abels2012,campillo-funollet2012}, in order for the model to be frame-invariant and not to introduce further nonlinearities, the momentum should be transported by the same mass-averaged velocity $\v u_{\rm mass}$.
This gives the following evolution equation for the momentum:
\begin{equation}
  \rho \pdt {\v u} + ( \v m \cdot \grad ) \v u - \div \v S + \grad p = \v K,
\end{equation}
where $\v K$ is a forcing term that will be determined by thermodynamic consistency, $\v S$ is a stress tensor to be decided, and $\v m = \rho \v u _{\rm mass} = \rho \v u + \v J_\rho $.

The electric field $\v E = - \grad V$ can be found through Gauss' law:
\begin{equation}
  \div \left[ \epsilon (\{ c_i \}) \v E \right] = \rho_e,
  \label{eq:pd_E}
\end{equation}
where the total charge is 
\begin{equation}
  \rho_e = \sum_i z_i c_i.
  \label{eq:rho_e}
\end{equation}
In Eq.\ \eqref{eq:pd_E} we have taken the permittivity, $\epsilon$, to be a function of the concentrations. 
This is motivated by, e.g, studies on aqueous NaCl solutions where it has been observed that the permittivity can be significantly reduced due to multibody effects \cite{hess2006}.
For simplicity we have dropped the weak dependence of permittivity on the electric fields \cite{booth1951} which for most purposes are insignificant \cite{pride1991}.
Now, using \eqref{eq:rho_e} and \eqref{eq:pd_c_i} we can write
\begin{equation}
  \pdt {\rho_e} + \div ( \rho_e \v u ) = - \sum_i \div (z_i \v J_{c_i}) ,
  = - \div \v J_e,
\end{equation}
where we have used that $\sum_i z_i R_i = 0$ due to charge conservation in a reaction, and defined $\v J_e = \sum_i z_i \v J_{c_i}$.
Using \eqref{eq:pd_E}, we find
\begin{equation}
  \pdt {\left[ \epsilon (\{ c_i \}) \v E \right]} + \rho_e \v u = - \v J_e .
\end{equation}
or
\begin{equation}
  \epsilon \pdt {\v E } = - \rho_e \v u - \v J_e -  \sum_i \v E \pd { \epsilon } {c_i} \pdt c_i.
\end{equation}

We can now define the following general free energy density $f$:
\begin{equation}
  f[\v u, \{c_k\}, \v E ] (\v x, t) = \frac{1}{2} \rho( \{c_k \} ) \v u ^2 + \sum_i M_i (\{ c_k \}) + \frac{1}{2} \epsilon (\{c_k\}) \v E^2 - \rho \, \v x \cdot {\v a_g}
  \label{eq:force_density}
\end{equation}
and thus the total energy density
\begin{equation}
  F = \intV f \, \dV .
\end{equation}
Now,
\begin{equation}
  \d{F}{t} = \intV \left[ \v u \cdot \rho \pdt \v u + \sum_i \pd f {c_i} \pdt c_i + \v E \cdot \epsilon \pdt \v E \right] \dV.
\end{equation}
Further,
\begin{equation}
  \pd f {c_i} = \left( \frac{\v u ^2}{2} - \v x \cdot {\v a_g} \right) \pd {\rho}{c_i} + \sum_k \pd {M_k} { c_i } + \frac{\v E^2}{2} \pd {\epsilon} {c_i} ,
\end{equation}
and hence
\begin{equation}
  \d{F}{t} = \intV \Bigg[ \v u \cdot \left( - ( \v m \cdot \grad ) \v u + \div \v S - \grad p + \v K \right) + \sum_i \pd f {c_i} \pdt c_i  
  - \v E \cdot \left( \rho_e \v u + \v J_e + \sum_i \v E \pd { \epsilon } {c_i} \pdt c_i \right) \Bigg] \, \dV  .
\end{equation}
Integrating by parts, using that all normal fluxes vanish at the boundary, we obtain
\begin{multline}
  \d F t = \intV \Bigg[ \sum_i ( g_i - z_i V) \pdt c_i 
 + \v u \cdot \v K - \rho_e \v E \cdot \v u - \v E \cdot \sum_i z_i \v J_{c_i}  \Bigg] \dV + \int_{\partial\Omega} \hat{\v n}  \cdot\v S \cdot \hat{\v t} u \, \diff \Gamma - \intV \v D \v u : \v S \, \dV
\end{multline}
where we have used that $\v u = \hat{\v t}  u $ on the boundary $\partial\Omega$, and defined the chemical potential
\begin{align}
  g_i &= \pd f {c_i} - \frac{1}{2} \v u ^2  \pd {\rho} {c_i} - \v E^2 \pd { \epsilon } {c_i} + z_i V\\
               &= - \v x \cdot {\v a_g} \pd {\rho}{c_i} + \sum_k \pd {M_k} { c_i } - \frac{\v E^2}{2} \pd {\epsilon} {c_i} + z_i V .
\end{align}
Now,
\begin{align}
  \sum_i \intV ( g_i - z_i V) \pdt c_i \dV &= - \sum_i \intV ( g_i - z_i V) ( \div ( c_i \v u  + \v J_{c_i} ) - R_i )\, \dV \\
                                    &= \sum_i \intV  ( \grad g_i + z_i \v E ) \cdot ( c_i \v u + \v J_{c_i}) \, \dV + \sum_i \intV g_i R_i \, \dV ,
\end{align}
such that
\begin{multline}
  \d F t = \intV \v u \cdot \left[ \v K + c_i \grad g_i \right] \dV + \int_{\partial\Omega} \hat{\v n} \cdot \v S \cdot \, \hat{\v t} u \, \diff \Gamma- \intV \v D \v u : \v S \dV
 + \sum_i \intV \v J_{c_i} \cdot \grad g_i \, \dV \\+ \sum_i \intV g_i R_i \, \dV .
 \label{eq:dFdt_general}
\end{multline}

To choose the fluxes according to Onsager's variational principle (as in Refs.\ \cite{abels2012,campillo-funollet2012}), we identify
\begin{equation}
  \v J_{c_i} = - K_i (\{c_k\}) \grad g_i,
\end{equation}
where $K_i \geq 0 $ are the mobilities.
Further, the viscosity tensor can be modelled with the Newtonian form,
\begin{equation}
  \v S = 2 \mu (\{ c_k \}) \v D \v u .
\end{equation}
Note that the viscosity $\mu \geq 0 $ can also depend on $\v D \v u$ to model non-Newtonian fluids, but we shall not consider that here.
To ensure that the contribution from the boundary integral be dissipative, we may choose the velocity boundary condition
\begin{equation}
  \hat{\v n} \cdot \v S \cdot \hat{\v t} = - \lambda \hat{\v t} u = - \lambda \v u,
\end{equation}
with $\lambda \geq 0$.
Finally, to minimize the dissipation we choose the forcing term according to
\begin{equation}
  \v K = - \sum_i c_i \grad g_i .
\end{equation}
The motivation for modelling the last term in \eqref{eq:dFdt_general} is given in \ref{sec:modelling_reaction_terms}. 

\begin{remark}
  In the definition of the force density in \cref{eq:force_density}, we could have used the mass-averaged velocity $\v u_{\rm mass}$ instead of the volume-averaged velocity $\v u$, which would have lead to a slightly different model.
  However, for simplicity, numerical purposes and to be consistent to two-phase flow models \cite{abels2012,campillo-funollet2012}, we have used the volume-averaged velocity $\v u$.
  For more in-depth discussions on the subtle difference between the mass-averaged and the volume-averaged velocity, we refer to the literature \cite{brenner2005navier,brenner2005kinematics,brenner2006fluid}.
\end{remark}
\section{Properties of the model}
\label{sec:properties}
In this section, we inspect some properties of the model presented in the preceding section.

\subsection{Evolution of ion concentration}
The first notable feature of the model is that the total ion concentration evolves only due to the reaction source term $R_i$:
\begin{equation}
  \d {} t \intV c_i \, \dV = \intV \pdt c_i \, \dV = - \intV \div \left( \v u c_i - K_i \grad g_i \right) \, \dV + \intV R_i \, \dV = \intV R_i \, \dV.
\end{equation}
When no reactions occur, the number of ions (integrated concentration) is conserved.

\subsection{Mass conservation}
The evolution of the density $\rho$ can be expressed by using Eqs.\ \eqref{eq:eos_rho_cj} and \eqref{eq:model_ci}:
\begin{equation}
  \pdt \rho = \sum_i \pd {\rho}{c_i} \pdt c_i = \sum_i \pd {\rho}{c_i} (- \div \v J_i + R_i) = - \div \v m,
  \label{eq:mass_conservation}
\end{equation}
where we have, as in the previous section, used the condition that a reaction can not change the density, i.e., $\sum_i \pd {\rho}{c_i} R_i = 0$.
Thus mass is conserved in the model:
\begin{equation}
  \d {} t \intV \rho \, \dV = \intV \pdt \rho \, \dV = - \intV \div \v m \, \dV = 0.
\end{equation}

\subsection{Free energy}
Associated with the above system we have the free energy
\begin{equation}
  F = \intV \left[ \frac{1}{2} \rho |\v u|^2 + \frac{1}{2}\epsilon |\grad V|^2 + \sum_i M_i - \rho \, \v a_g \cdot \v x \right] \dV,
\end{equation}
where the first term represents the kinetic energy, the second the electric field energy, the third term the chemical energy,  and the last term the gravitational energy.

Energy-stability imparts that, in the absence of external driving forces, the total free energy $F$ in the closed domain does not increase in time; that is, $\diff F / \diff t \leq 0$.
In order to inspect whether the model respects this, we are now interested in an explicit expression for the evolution of the free energy in time, i.e.~$\diff F / \diff t$.
We therefore decompose the free energy $F$ into the following contributions:
\begin{equation}
  F = F_{\v u} + F_V + \sum_i F_{c_i} + F_g,
  \label{eq:F_dec}
\end{equation}
where
\begin{subequations}
\begin{align}
  F_{\v u} &= \intV \frac{1}{2} \rho |\v u|^2 \, \dV, & &\text{(kinetic energy)} \label{eq:def_Fu}\\
  F_V &= \intV \frac{1}{2}\epsilon |\grad V|^2 \, \dV,  & &\text{(electric field energy)} \label{eq:def_FV} \\
  F_{c_i} &= \intV M_i \, \dV,  & &\text{(chemical energy)} \label{eq:def_Fci}\\
  F_g &= -\intV \rho \, \v a_g \cdot \v x \,\dV.  & & \text{(gravitational energy)} \label{eq:def_Fg}
\end{align}
\label{eq:F_x}
\end{subequations}
Now, we seek the temporal evolution of each term in order to find the temporal evolution of the sum of them.

\begin{itemize}
\item \textbf{The kinetic energy:} Differentiating Eq.\ \eqref{eq:def_Fu} with respect to time, integrating by parts, and using \eqref{eq:model_NS1}, \eqref{eq:model_NS2}, and the boundary conditions \eqref{eq:unBC}, \eqref{eq:utBC} and \eqref{eq:gBC}, we obtain
\begin{align}
  \d {F_{\v u}} t &= \intV \pdt \left[ \frac{1}{2} \rho |\v u|^2 \right] \dV \nonumber\\
  &= ( \v u , \rho \pdt \v u ) + \left( \frac{1}{2} |\v u|^2 , \pdt \rho \right) \nonumber\\ 
  &= ( \v u, \div ( 2 \mu \v D \v u ) + \grad p - \sum_i c_i \grad g_i  
  ) \nonumber\\
  &= - \norm{\sqrt{2 \mu} \v D \v u}^2 - \norm{\sqrt{\lambda} \v u}^2_{\partial\Omega} 
  - \sum_i (\v u, c_i \grad g_i ) \nonumber\\
  &= - \norm{\sqrt{2 \mu} \v D \v u}^2 - \norm{\sqrt{\lambda} \v u}^2_{\partial\Omega}
  - \sum_i \left[ ( g_i, \pdt c_i) + \norm{ \sqrt{ K_i} \grad g_i }^2 - (g_i, R_i) \right],
    \label{eq:dFudt}
\end{align}
where we have used the fact that $K_i$ is non-negative.
Note that for the standard Nernst--Planck assumption $K_i = D_i c_i$, this relies on having established a non-negativity result for $c_i$, or alternatively using a regularised $K_i = D_i \max(c_i, 0)$.
For the standard Nernst--Planck equations, a non-negativity result for $c_i$ was provided in \cite{schmuck2009}.
In \ref{sec:proof-non-negative}, we outline a simple proof for the model considered here.

\item \textbf{The electric field energy:} Differentiating Eq.\ \eqref{eq:def_FV}, integrating by parts and using Eq.\ \eqref{eq:model_V} and the boundary conditions \eqref{eq:VBC}, we obtain
\begin{align}
  \d {F_V} t &= \d {} t \int_\Omega \frac{1}{2} \epsilon |\grad V|^2 \, \diff \Omega \nonumber\\
  &= (  \grad V, \epsilon \pdt \grad V) + \left( \frac{1}{2} |\grad V|^2 , \pdt \epsilon \right) \nonumber\\
  &= (  \grad V, \pdt (\epsilon \grad V)- \grad V \pdt \epsilon ) + \left( \frac{1}{2} |\grad V|^2 , \pdt \epsilon \right) \nonumber\\
  &= (  \grad V, \pdt (\epsilon \grad V) ) - \left( \frac{1}{2} |\grad V|^2 , \pdt \epsilon \right) \nonumber\\
  &= - ( V, \pdt \div (\epsilon \grad V) ) - \sum_i \left( \frac{1}{2} |\grad V|^2 , \pd {\epsilon}{c_i} \pdt c_i \right) \nonumber\\
  &= \sum_i ( z_i V - \frac{1}{2} |\grad V|^2 \pd {\epsilon}{c_i}, \pdt c_i ) 
                \label{eq:dFVdt}
\end{align}

\item \textbf{The chemical energy:}
  Differentiating Eq.\ \eqref{eq:def_Fci} with respect to time and using the assumption that $M_i$ only depends on the set of concentrations $\{ c_k \}$, we can write
\begin{align}
  \d {F_{c_i}} t &= \int_\Omega \pdt M_i \, \diff \Omega = \sum_j \int_\Omega \pd {M_i} {c_j} \pdt c_j \, \diff \Omega .
\end{align}

\item \textbf{The gravitational energy:} By differentiation of Eq.\ \eqref{eq:def_Fg} and using Eq.\ \eqref{eq:eos_rho_cj}, we obtain
\begin{align}
  \d {F_g} t &= -\intV \pdt \rho \, \v a_g \cdot \v x \, \dV \nonumber \\ 
   &=  - \sum_i \left( \pd {\rho}{c_i} \v a_g \cdot \v x , \pdt c_i \right) .
     \label{eq:dFgdt}
\end{align}

\end{itemize}

Using \cref{eq:dFudt,eq:dFVdt,eq:dFgdt,eq:F_dec} and the definition of $g_i$ in Eq.\ \eqref{eq:model_gi} we obtain:
\begin{align}
  \d F t = - \norm{\sqrt{2 \mu } \v D \v u}^2- \norm{\sqrt{\lambda} \v u}^2_{\partial\Omega} - \sum_i \norm{ \sqrt{ K_i} \grad g_i }^2 
  + \sum_i \left( g_i, R_i \right).
  \label{eq:ineq_F_cont}
\end{align}
Clearly, the three first terms on the right hand side of Eq.\ \eqref{eq:ineq_F_cont} are negative.
Thus, what remains is to model the reaction terms $R_i$ in such a way that the last term is also negative.
Inserting Eq.\ \eqref{eq:reaction_R_text} (as discussed in \ref{sec:modelling_reaction_terms}) into the last term leads to
\begin{align}
  \sum_i (g_i, R_i) &= -  \sum_i (g_i, \sum_m \nu_{m,i} \mathcal{C}_m \cdot \sum_j \nu_{m,j} g_j) \nonumber\\
                    &= - \sum_m \mathcal{C}_m \int_\Omega \left( \sum_i \nu_{m,i} g_i\right)^2 \dV.
                      \label{eq:giRi}
\end{align}

The sought evolution equation is now readily available.
In particular, we obtain from Eqs.\ \eqref{eq:ineq_F_cont} and \eqref{eq:giRi} the free energy evolution
\begin{equation}
    \d F t = - \norm{\sqrt{2 \mu } \v D \v u}^2- \norm{\sqrt{\lambda} \v u}^2_{\partial\Omega} - \sum_i \norm{ \sqrt{ K_i} \grad g_i }^2 
    - \sum_m \mathcal{C}_m \intV \left( \sum_i \nu_{m,i} g_i \right)^2 \, \dV \leq 0.
    \label{eq:evolution_inequality}
\end{equation}
Hence the free energy is decaying in time --- i.e.\ the model is dissipative. This is an important property, as it guarantees that, in the absence of external driving forces, the system at all instances does not produce energy, i.e.~it evolves towards a state of lower energy.
Hence, a proper time discretization scheme should also have this property, in order to avoid spurious energy blow-up.

\subsection{Modelling choices}
\label{sec:modelling_choices}
Note that we will not attempt to quantitatively model the reaction function $\mathcal{C}_m$ (apart from the example considered in \ref{sec:modelling_reaction_terms}).
This will in general require more detailed or phenomenological modelling of the particular chemical reaction $m$.

In the remainder of this article, we will for concreteness consider the chemical energy functions
\begin{equation}
  M_i (\{ c_k \}) = \alpha ( c_i ) + \beta_{i} c_i,
\end{equation}
where $\beta_{i}$ are constants.
The role of $\beta_{i}$ is to energetically penalize (or promote) the presence of a species $c_i$ in comparison to other species.
Hence, the set $\{\beta_{i}\}$ should fix a (chemical) equilibrium state of the system.
In particular, for the common form $\alpha(c_i) = c_i (\ln c_i-1)$, we can write
\begin{equation}
  \beta_i = - \ln c_i^0,
\end{equation}
where $c_i^0$ is a reference concentration which defines the equilibrium concentration of species $c_i$.
(See \ref{sec:modelling_reaction_terms} for its relation to the solubility product.)

The derivative of $M_i$ that enters into the model can be expressed by
\begin{equation}
  \pd {M_i} {c_j} = \alpha'( c_j ) \delta_{ij} + \beta_{j} \delta_{ij},
\end{equation}
where $\delta_{ij}$ is the Kronecker delta function, and we have used the short-hand derivative $\alpha'( c_j) = \pdinl{\alpha} {c_j}$. 
Note that since the $\beta_{i}$ are constant, they will not affect the system through the chemical diffusive fluxes ($\propto K_i \grad g_i$), but will enter in the reaction term $R_i$.

Further, we will consider only permittivities which can be written in the form
\begin{equation}
  \epsilon (\{ c_j \}) = \epsilon_0 + \sum_{j=1}^N \epsilon_j ( c_j ),
  \label{eq:def_epsilon}
\end{equation}
where, in particular, no cross terms are present.
Here, $\epsilon_0$ is not the vacuum permittivity, but the permittivity of the background fluid.
Note that on physical grounds $\epsilon > 0$ (in particular, the vacuum permittivity is an absolute lower bound).
The formulation \eqref{eq:def_epsilon} is consistent, e.g., with the empirical relation found in simulations by \citet{hess2006} for a NaCl solution, where a relation $1/\epsilon(c) \propto 1 + k c$ ($k$ is a constant) was reported.
\citet{gavish2016} found a somewhat more complicated, but qualitatively similar, functional relation, valid for a variety of salts and concentrations.

\section{Energy-stable time discretization}
\label{sec:numerics}
We will in the forthcoming consider schemes that are finite difference in time, and finite element in space.
We present schemes to simulate the general model for single-phase electrohydrodynamics which was presented in the previous section.
In this section, we will first present the schemes and afterwards the appropriate variational form which is used in the finite element spatial discretization.
As our main focus is on the temporal discretization of the model, we consider a continuous-space (and discrete-time) finite-element formulation rather than a fully discrete spatial discretization.
To this end, for the velocity components we define the function space $\mathcal V$ as
\begin{equation}
  \mathcal V = \{ v \in H^1(\Omega): v = 0 \ \textrm{on} \ \partial \Omega \}
\end{equation}
where $H^1(\Omega)$ is the Sobolev space containing functions $f$ such that  $f^2$ and $|\grad f|^2$ have finite integrals over $\Omega$.
To ease notation in the following, we will for the remaining, scalar fields use the spaces $\mathcal X$ which we define as simply $\mathcal V$ without the boundary restrictions, i.e.\ $\mathcal X = H^1(\Omega)$.

Since a central point of this article is to construct energy-stable schemes, it is necessary to define what this concept imparts.
Analogous to previous work in the literature, e.g.\ \cite{minjeaud2013,shen2014,shen2015,guillen-gonzalez2014}, we state the following definition.
\begin{defn}[Energy stability]
  Recall that the model \eqref{eq:model_NS1}--\eqref{eq:model_V} is associated with the free energy functional $F$, stated in Eqs.\ \eqref{eq:F_dec} and \eqref{eq:F_x}.
  The temporal derivative of $F$ satisifies the inequality
\begin{equation}
  \d{F}{t} \leq 0,
  \label{eq:dFdtleq0}
\end{equation}
as stated in Eq.\ \eqref{eq:evolution_inequality}, meaning that free energy is dissipated.

Now, let $F^k$ be the discrete counterpart of $F$ -- that is, an approximation to $F$ at time $t^k$ (at time step $k$).
If \emph{a temporal discretization scheme} unconditionally satisfies the discrete counterpart of Eq.\ \eqref{eq:dFdtleq0}, that is,
\begin{equation}
  \dpdt F^k \leq 0,
\end{equation}
where we have used Eq.\ \eqref{eq:def_dpdt},
\emph{then that scheme is said to be energy stable}.
\end{defn}

Note that for a scheme to be admissible, it must obviously also approximate the underlying model to the required order in the time step size $\tau$.
As is remains unclear how to construct temporally second-order schemes that are simultaneously energy stable, we consider only first-order schemes in the present work.

\subsection{Decoupled schemes}
We will in this paper adopt a strategy known from simulating, e.g., two-phase flow.
It is beneficial to split the problem in a hydrodynamical step and an electrochemical step, since it is in general harder both to effectively precondition and to solve the coupled system.
On the other hand, there exists approaches for efficient solution of the separate subproblems; i.e., for the PNP system (for the electrochemistry) and for the NS system (for the hydrodynamics).
The decoupling strategy may also enable the construction of linear schemes, instead of non-linear, thus possibly saving the excess computatation related to nonlinear iterations.

The main advantage of the schemes presented here is that the computation of the electrochemical problem is decoupled from the hydrodynamic problem, while we are still able to guarantee the energy dissipation associated with the physical problem.

Hence, we shall now consider schemes which employ a divide-and-conquer strategy, with two subproblems to be solved \emph{sequentially} at each time step $k$:
\begin{enumerate}
\item \textbf{Electrochemistry:} Using information from the previous time step $k-1$, i.e., $\{\v u^{k-1}$, $p^{k-1}$, $c_1^{k-1}$, $\ldots$, $c_N^{k-1}$, $V^{k-1}\}$, obtain a numerical approximation of the primary electrochemical variables, i.e.~$\{c_1^k$, $\ldots$, $c_N^k$, $V^k\}$ at the present time step $k$.
\item \textbf{Hydrodynamics:} Using the newly updated electrochemical variables $\{c_1^k$, $\ldots$, $c_N^k$, $V^k\}$ and hydrodynamic variables $\{\v u^{k-1}$, $p^{k-1}\}$ from the previous time step $k-1$, obtain an approximation of the primary hydrodynamical variables, i.e.~$\{\v u^k$, $p^k\}$ at the present time step $k$.
\end{enumerate}


\subsection{Strategy for the electrochemistry step}
\paragraph{Scheme}
Suppose $\{ \v u^{k-1}, p^{k-1}, c_1^{k-1}, \ldots, c_N^{k-1}, V^{k-1} \}$ are given. Now, to obtain $\{ c_1^k, \ldots, c_N^k, V^k \}$, solve
\begin{subequations}
\begin{equation}
  \dpdt{c_i^k}
  - \div \left( \v u^* \tilde c_i \right)
  - \div \left( \tilde K_i \grad g_i^k \right) = \tilde R_i, \quad \textrm{for} \quad i \in [1, N],
  \label{eq:single_c1_scheme}
\end{equation}
\begin{equation}
  \div \left( \epsilon^k \grad V^k \right) = \sum_i z_i c_i^k ,
  \label{eq:single_V_scheme}
\end{equation}
where
\begin{equation}
  g_i^k = \tilde{\alpha}' 
    + \beta_i 
    + z_i V^k - \frac{1}{2} |\grad{ V^k}|^2 \tilde{\epsilon}_i' - \pd{\rho}{c_i} \v x \cdot \v a_g .
  \label{eq:single_c2_scheme}
\end{equation}
\end{subequations}
Here, $\tilde{\alpha}'(c_i^k, c_i^{k-1})$ is a numerical approximation to $\alpha'(\xi^k)$, where $\min(c_i^k, c_i^{k-1}) \leq \xi^k \leq \max(c_i^k, c_i^{k-1})$.
Further, $\tilde K_i (c_i^k, c_i^{k-1}) \geq 0$ approximates $K_i$, $\tilde c_i(c_i^k, c_i^{k-1})$ is an approximation to $c_i$, and $\tilde R_i$ is an approximation to $R_i$.
We use the generally formulated terms $\tilde K_i$, $\tilde c_i$, and $\tilde R_i$, to keep the analysis as general as possible, and valid for both explicit (dependent on $c_i^{k-1}$) and implicit (dependent on $c_i^k$) discretizations.
Later, we will discuss concrete examples of the stated terms.

Moreover,
\begin{equation}
\tilde\epsilon_i' ( c_i^k,  c_i^{k-1}) = \begin{cases}
  \frac{\epsilon_i( c_i^k)-\epsilon_i(c_i^{k-1} )}{c_i^k-c_i^{k-1}} & \textrm{for} \quad c_i^k \neq c_i^{k-1},\\
  \pd{\epsilon_i} {c_i} (c_i^{k-1}), & \textrm{for} \quad c_i^k = c_i^{k-1},
\end{cases}
\end{equation}
is an approximation to $\pdinl {\epsilon_i} {c_i}$.
Recall also that $\pdinl {\rho}{c_i}$ is a constant.

The following boundary conditions are enforced on the boundary $\partial\Omega$
\begin{subequations}
\begin{gather}
  \hat{\v n} \cdot \tilde K_i \grad g_i^k = 0, \\
  \hat{\v n} \cdot \epsilon^k \grad V^k = \sigma_e \quad \textrm{or} \quad V^k =0.
\end{gather}
\end{subequations}

In \cref{eq:single_c1_scheme} we have used the definition:
\begin{equation}
  \v u^* = \v u^{k-1} - \frac{\tau}{\rho^{k-1}} \sum_i \tilde c_i \grad g_i^k ,
  \label{eq:single_ustar}
\end{equation}
which is a forward-projection of the velocity based on the chemical fluxes, and introduces a first-order error in $\tau$.
This projection is a key ingredient to obtaining energy-stability and is inspired by schemes for two-phase flow \cite{minjeaud2013,shen2015,metzger2015}.
Note that when the system approaches equilibrium, the second term, which is already close to equilibrium, vanishes.

In practice, the inverse density $1/\rho^{k-1}$, which enters in the second term of \cref{eq:single_ustar}, must be approximated by finite element functions.
On P$_1$ elements, it can be straightforwardly approximated by pointwise interpolation on the mesh nodes, as done herein.
Other interpolation procedures are possible, but we have not considered these here, as our focus is primarily on temporal rather than spatial discretization strategies.

\paragraph{Variational form}
A variational form of \cref{eq:single_c1_scheme,eq:single_c2_scheme,eq:single_V_scheme} can be written as the following.

Find $(c_1^k, \ldots, c_N^k, g_1^k, \ldots g_N^k, V^k) \in \mathcal X^N \crossproduct \mathcal X^N \crossproduct \mathcal X$, such that for all $(b_1^k, \ldots, b_N^k, h_1^k, \ldots h_N^k, U^k) \in \mathcal X^N \crossproduct \mathcal X^N \crossproduct \mathcal X$, we have
\begin{subequations}
\begin{equation}
  \left(\dpdt{c_i^k}, b_i \right)
  - \left( \v u^* \tilde c_i , \grad b_i \right)
  + \left( \tilde K_i \grad g_i^k , \grad b_i \right)  = \left( \tilde R_i, b_i \right),
  \label{eq:single_c1}
\end{equation}
\begin{equation}
  \left( g_i^k, h \right) = \left( \tilde{\alpha}' 
    + \beta_i 
    + z_i V^k - \frac{1}{2} |\grad{ V^k}|^2 \tilde{\epsilon}_i' - \pd{\rho}{c_i} \v x \cdot \v a_g, h\right)
  \label{eq:single_c2}
\end{equation}
\begin{equation}
  \left( \epsilon^k \grad V^k, \grad U \right) - \int_{\partial\Omega} \sigma_e U \, \diff \Gamma = \sum_{i=1}^N \left( z_i c_i^k, U \right).
  \label{eq:single_V}
\end{equation}
\end{subequations}

\subsubsection{Free energy evolution}
\begin{property}
  \label{lemma:EC}
  For the electrochemical step, the following inequality holds:
  \begin{equation}
    \dpdt F_{\rm EC}^k
    \leq
    \sum_i \left( \v u^*, \tilde c_i \grad g_i^k \right) 
    - \frac{1}{\tau} \sum_i \Delta F_{c_i}^k
    + \sum_i \left( \tilde R_i , g_i^k \right)
    .
    \label{eq:ineq_EC}
  \end{equation}
  Here, the discrete electrochemical energy is given by
  \begin{equation}
    F_{\rm EC}^k = \sum_i F_{c_i}^k + F_V^k + F_g^k,
  \end{equation}
  and we have defined
  \begin{equation}
     \Delta F^k_{c_i} = \tau \left( \tilde{\alpha}'(c_i^k, c_i^{k-1}) + \beta_i, \dpdt c_i^k \right) - F_{c_i}^k + F_{c_i}^{k-1},
    \label{eq:deltaFdef}
  \end{equation}
  which represents an approximation error in the free energy introduced by the numerical approximation $\tilde{\alpha'}(c_i^k, c_i^{k-1})$ to $\alpha'(c)$.
  If $\Delta F^k_{c_i} \geq 0$, it can be seen as an excess free energy.
\end{property}

\begin{proof}
  By testing \cref{eq:single_c1} with $b_i = g_i^k$, we get:
  \begin{equation}
    \left( \dpdt{c_i^k}, g_i^k \right)
    - \left( \v u^* \tilde c_i, \grad g_i^k \right)
    = - \norm{ \sqrt{ \tilde K_i } \grad g_i^k }^2
    + \left( \tilde R_i , g_i^k \right)
    ,
    \label{eq:single_c1-2}
  \end{equation}
  and further, testing \cref{eq:single_c2} with $h = \dpdt{c_i^k}$, we obtain:
  \begin{align}
    \left( g_i^k, \dpdt{c_i^k} \right)
    &= \frac{1}{\tau} \left( \tilde{\alpha}'(c_i^k, c_i^{k-1}) + \beta_i + z_i V^k  - \frac{1}{2} |\grad{V^k}|^2 \tilde{\epsilon}_{i}' - \pd{\rho}{c_i} \v x \cdot \v a_g, c_i^k - c_i^{k-1} \right) \nonumber \\
      &= \dpdt F_{c_i}^k + \frac{\Delta F_{c_i}^k}{\tau} + \left( z_i V^k, \dpdt{c_i^k} \right)
        - \left( \frac{1}{2} |\grad{V^k}|^2 , \dpdt{\epsilon_i^k} \right)
        - \left( \pd{\rho}{c_i} \v x \cdot \v a_g, \dpdt{c_i^k} \right)
        ,
        \label{eq:single_c2-2}
  \end{align}
  where we have introduced the splitting \eqref{eq:deltaFdef} and the shorthand definition of the discrete total chemical energy:
  \begin{equation}
    F_{c_i}^k = \intV \left[ \alpha(c_i^{k}) + \beta_i c_i^k \right]\, \dV.
  \end{equation}

  By defining the shorthand discrete gravitational energy,
  \begin{equation}
    F_g^k = - \intV \rho^k \, \v x \cdot \v a_g \, \dV,
  \end{equation}
  where $\rho^k = \rho(\{c_i^k \})$, we find that the sum over the phases in the last term in Eq.\ \eqref{eq:single_c2-2} becomes
  \begin{equation}
    \sum_i \left( \pd{\rho}{c_i} \v x \cdot \v a_g, \dpdt{c_i^k} \right) = \left( \v x \cdot \v a_g, \dpdt{ \left[ \rho_0 + \sum_i \pd{\rho}{c_i} c_i^k \right]} \right) = \left( \v x \cdot \v a_g, \dpdt \rho^k \right) = - \dpdt F_g^k .
    \label{eq:single_g}
  \end{equation}

  We also define the discrete electric energy by
  \begin{equation}
    F_V^k = \intV  \frac{1}{2} \epsilon^k |\grad V^k|^2 \, \dV .
  \end{equation}
  
  Now, testing \cref{eq:single_V} with $U=V^k$ yields:
  \begin{equation}
    \left( \epsilon^k \grad V^k, \grad V^k \right) - \int_{\partial\Omega} \sigma_e V^k \, \diff \Gamma= \sum_i \left( z_i c_i^k, V^k \right) .
    \label{eq:single_V-2}
  \end{equation}
  Considering \cref{eq:single_V} with $k \to k-1$, and testing it with $U=V^k$, yields:
  \begin{equation}
    \left( \epsilon^{k-1} \grad V^{k-1}, \grad V^{k}\right) - \int_{\partial\Omega} \sigma_e V^k \, \diff \Gamma = \sum_i \left( z_i c_i^{k-1}, V^k \right) .
    \label{eq:single_V-3}
  \end{equation}
  Subtracting \cref{eq:single_V-3} from \cref{eq:single_V-2} and dividing by $\tau$ gives
  \begin{align}
    \sum_i z_i \left( \dpdt{c_i^k}, V^k \right)
    &= \frac{1}{\tau} \left( \grad V^k, \epsilon^k \grad V^k - \epsilon^{k-1} \grad V^{k-1} \right) \nonumber \\
    &= \left( \frac{\v E^k + \v E^{k-1}}{2\tau} + \frac{\v E^k - \v E^{k-1}}{2\tau}, \epsilon^k \left(\v E^k - \v E^{k-1} \right) \right) + \left( \v E^k, \dpdt \epsilon^{k} \v E^{k-1} \right) \nonumber\\
    &= \frac{1}{2\tau} \left( \epsilon^k, |\v E^k|^2 - |\v E^{k-1}|^2 \right) + \frac{1}{2\tau} \norm{\sqrt{\epsilon^k} \left(\v E^k - \v E^{k-1} \right) }^2  + \left( \v E^k, \dpdt \epsilon^{k} \v E^{k-1} \right)\nonumber \\
    &= \dpdt F_V^k 
      + \frac{1}{2\tau} \norm{\sqrt{\epsilon^{k-1}} \left(\grad V^k - \grad V^{k-1} \right) }^2
      + \frac{1}{2} \left( \dpdt \epsilon^k, 
      |\grad V^k|^2 
      \right)
      \label{eq:single_V-4}                                                   
  \end{align}
  Now, combining \cref{eq:single_c1-2,eq:single_c2-2,eq:single_g,eq:single_V-4}, we obtain
  \begin{multline}
     \sum_i \dpdt F_{c_i}^k
    + \dpdt F_V^k
    + \dpdt F_g^k
    =
    - \frac{1}{\tau} \sum_i \Delta F_{c_i}^k
    + \sum_i \left( \v u^*, \tilde c_i \grad g_i^k \right) \\
    - \sum_i \norm{ \sqrt{ \tilde K_i} \grad g_i^k }^2
    - \frac{1}{2\tau} \norm{\sqrt{\epsilon^{k-1}} \left(\grad V^k - \grad V^{k-1} \right) }^2
    + \sum_i \left( \tilde R_i , g_i^k \right)
    .
    \label{eq:single_part4}
  \end{multline}
  which yields \cref{eq:ineq_EC} and thus completes the proof.
  \end{proof}


\subsection{Strategies for the hydrodynamic step}
For the hydrodynamic step, we can consider either the standard coupled approach, which is to solve the velocity and pressure simultaneously at each time step, or an approach which decouples the velocity and pressure at each step.
We shall denote the former as Scheme I and the latter as Scheme II.

\subsubsection{Scheme I: Coupled hydrodynamics}
\paragraph{Scheme}
The first scheme can be written in variational form as the following.
Suppose that $\{ \v u^{k-1},$ $p^{k-1},$ $c_1^{k-1},$ $\ldots,$ $c_N^{k-1},$ $c_1^{k},$ $\ldots,$ $c_N^{k},$ $g_1^{k},$ $\ldots,$ $g_N^{k}\}$ are given.
Now, in order to obtain $\{ \v u^k, p^k \}$, we solve
\begin{subequations}
\begin{equation}
  \rho^{k-1} \dpdt {\v u^k}
   + (\v m^{k-1} \cdot \grad ) \v u^{k}
  - \div \left( 2 \mu^k \v D \v u^{k} \right)
  + \grad p^k
  + \frac{1}{2} \v u^k \left( \dpdt \rho^k + \div \v m^{k-1} \right)
  = - \sum_i \tilde c_i \grad g_i^k, 
  \label{eq:single_NS1_scheme}
\end{equation}
\begin{equation}
  \div \v u^{k} = 0.
  \label{eq:single_NS2_scheme}
\end{equation}
\end{subequations}
Note that the last two terms on the left hand side of Eq.\ \eqref{eq:single_NS1} are an approximation to the mass conservation equation \eqref{eq:mass_conservation}, i.e., $\pdt \rho + \div \v m = 0$.
The incorporation of these terms is a standard way of satisfying the discrete energy law at each time step (see e.g.\ \cite{shen2015}).
The equations \eqref{eq:single_NS1_scheme} and \eqref{eq:single_NS2_scheme} are solved in combinaton with the Navier slip condition (cf.\ Eq.\ \eqref{eq:NavierSlipBC}),
\begin{subequations}
  \begin{align}
    \v u^k \cdot \hat{\v n} &= 0, \label{eq:unBCk} \\
    \left[ \lambda^k \v u^k + 2\mu^k \v D \v u^k \cdot \hat{\v n}\right]  \cdot \hat{\v t} &= \v 0, \label{eq:utBCk}
  \end{align}
  \label{eq:NavierSlipBCk}
\end{subequations}
on the boundary $\partial\Omega$.

\paragraph{Variational form}
Find $( \v u^{k}, p^k ) \in \mathcal V^d \crossproduct \mathcal X$ such that for all $(\v v, q) \in \mathcal V^d \crossproduct \mathcal X$,
\begin{subequations}
\begin{multline}
  \left( \rho^{k-1} \dpdt {\v u^k}, \v v \right)
  + \left( (\v m^{k-1} \cdot \grad ) \v u^{k} , \v v \right)
  + \int_{\partial\Omega} \lambda^k \v u^k \cdot \v v \, \diff \Gamma
  + \left( 2 \mu^k \v D \v u^{k}, \v D \v v \right)
  - \left(p^k, \div \v v \right)
  \\
  + \frac{1}{2} \left(  \v u^k \dpdt \rho^k, \v v \right) - \frac{1}{2} \left( \v m^{k-1} ,  \grad (\v u^k \cdot \v v) \right)
  = - \sum_i \left( \tilde c_i \grad g_i^k, \v v \right), 
  \label{eq:single_NS1}
\end{multline}
\begin{equation}
  \left( q, \div \v u^{k} \right) = 0,
  \label{eq:single_NS2}
\end{equation}
\end{subequations}
with the Dirichlet no-penetration boundary condition \eqref{eq:unBCk} on $\partial\Omega$.

\subsubsection{Scheme II: Fractional-step hydrodynamics}
Instead of solving for velocity and pressure in a coupled manner, we may use a projection method to decouple the velocity computation from the pressure.
Such a scheme describing the somewhat similar equations of two-phase flow, was already proposed by, e.g., \citet{shen2015}.

\paragraph{Scheme}
In the spirit of the latter reference, the scheme is given by the following.
Suppose that $\{ \v u^{k-1},$ $p^{k-1},$ $c_1^{k-1},$ $\ldots,$ $c_N^{k-1},$ $c_1^{k},$ $\ldots,$ $c_N^{k},$ $g_1^{k},$ $\ldots,$ $g_N^{k}\}$ are given.
\begin{itemize}
\item \textbf{Tentative velocity step:} To obtain the intermediate velocity $\tilde {\v u}^k$, solve
\begin{multline}
  \rho^{k-1} \frac{\tilde{\v u}^{k}-\v u^{k-1}}{\tau} 
  + (\v m^{k-1} \cdot \grad ) \tilde{\v u}^{k}
  - \div \left( 2 \mu^k \v D \tilde{\v u}^{k} \right)
  + \grad p^{k-1} \\
  + \frac{1}{2} \tilde{\v u}^k \left( \dpdt \rho^k + \div \v m^{k-1} \right)
  = - \sum_i c_i^{k-1} \grad g_i^k,
  \label{eq:split_u_1_scheme}
\end{multline}
with 
the Navier slip boundary condition
\begin{subequations}
  \begin{align}
    \tilde{\v u}^k \cdot \hat{\v n} &= 0, \label{eq:unBCktilde} \\
    \left[ \lambda^k \tilde{\v u}^k + 2\mu^k \v D \tilde{\v u}^k \cdot \hat{\v n}\right]  \cdot \hat{\v t} &= \v 0, \label{eq:utBCktilde}
  \end{align}
  \label{eq:NavierSlipBCktilde}
\end{subequations}
on $\partial\Omega$.
\item \textbf{Pressure correction step:}
To obtain the corrected pressure $p^k$, solve
\begin{align}
  \laplacian (p^k - p^{k-1}) = \frac{\rho_0}{\tau} \div \tilde{\v u}^k,
  \label{eq:split_p_scheme}
\end{align}
with the artificial Neumann condition $\v n \cdot \grad (p^k - p^{k-1}) = 0$.
Note that this introduces an $O(\tau)$ error at the boundary.
\item \textbf{Velocity correction step:}
To obtain the final velocity $\v u^k$, solve
\begin{align}
  \rho^{k} \frac{\v u^k -\tilde{\v u}^k}{\tau} = - \grad \left( p^k - p^{k-1} \right),
  \label{eq:split_u_2_scheme}
\end{align}
with the Dirichlet boundary condition $\hat{\v n} \cdot \v u^k = 0$, which supresses the error from the Neumann condition above.
\end{itemize}

Together with the analysis in the previous section, this constitutes a scheme which is decoupled between the three parts electrostatics, velocity and pressure.
Therefore, it is significantly easier to solve than the fully coupled problem, and easier than solving for only velocity and pressure in a coupled manner.

\paragraph{Variational form}
\begin{itemize}
\item \textbf{Tentative velocity step:}
  Find $\tilde{\v u}^k \in \mathcal V^d$ such that for all $\v v \in \mathcal V^d$,
\begin{multline}
  \left(\rho^{k-1} \frac{\tilde{\v u}^{k}-\v u^{k-1}}{\tau}, \v v\right) 
  + \left( (\v m^{k-1} \cdot \grad ) \tilde{\v u}^{k} , \v v \right)
  + \int_{\partial\Omega} \lambda^k \tilde{\v u}^k \cdot \v v \, \diff \Gamma
  + \left( 2 \mu^k \v D \tilde{\v u}^{k}, \v D \v v \right)
  - \left(p^{k-1}, \div \v v \right) \\
  + \frac{1}{2} \left(  \tilde{\v u}^k \dpdt \rho^k, \v v \right) - \frac{1}{2} \left( \v m^{k-1} ,  \grad (\tilde{\v u}^k \cdot \v v) \right)
  = - \sum_i \left( c_i^{k-1} \grad g_i^k, \v v \right),
  \label{eq:split_u_1}
\end{multline}
with the Dirichlet boundary condition 
\eqref{eq:unBCktilde} 
on $\partial\Omega$.
\item \textbf{Pressure correction step:}
Find $p^k \in \mathcal X$ such that for all $q \in \mathcal X$, we have
\begin{align}
  \left( \frac{1}{\rho_0} \grad (p^k - p^{k-1}) , \grad q \right) = - \frac{1}{\tau} \left( \div \tilde{\v u}^k, q \right).
  \label{eq:split_p}
\end{align}
\item \textbf{Velocity correction step:}
Then, find $\v u^k \in \mathcal V^d$ such that for all $\v v \in \mathcal V^d$,
\begin{align}
  \left( \rho^{k} \frac{\v u^k -\tilde{\v u}^k}{\tau}, \v v \right) = \left( p^k - p^{k-1} , \div \v v \right),
  \label{eq:split_u_2}
\end{align}
which we solve by explicitly imposing the Dirichlet boundary condition $\hat{\v n} \cdot \v u^k = 0$. 
\end{itemize}

Note that using $\v v = (\rho^k)^{-1} \grad q$ in \cref{eq:split_u_2} yields, in combination with \cref{eq:split_p}
\begin{equation}
  \left( \div \v u^k, q \right) = \tau^2 \left( \left(\frac{1}{\rho^k}-\frac{1}{\rho_0}\right) \grad (\dpdt p^k) , \grad q \right) 
  ,
  \label{eq:divfree}
\end{equation}
i.e., that the fractional-step scheme introduces a weak compressibility of order $O(\tau^2)$, which becomes increasingly small when $\rho^k \simeq \rho_0$.
When the density does not vary with concentration, $\rho^k = \rho_0$ and the final velocity field $\v u^k$ is divergence free.

\begin{remark}
  With a slight reformulation of the variational problem, we can simplify the computation of the velocity steps $\tilde{\v u}^k$ and $\v u^k$, by solving for each of the components successively, since in the decoupled approach none of the components $\tilde{u}_j^k$ and $u_j^k$, $j \in \{1, \ldots, d\}$ of $\tilde{\v u}^k$ and $\v u^k$, respectively, depend on the other components.
  This simplification is fairly commonplace \cite{mortensen2015}.
  We shall leave this technical detail for further work.
\end{remark}

\subsubsection{Free energy evolution}
Now we set out to show that a free energy inequality is satisfied for a discrete time update.

\begin{property}
  \label{lemma:NS}
  For the hydrodynamic step, the following inequality holds: 
  \begin{equation}
    \dpdt F_{\rm NS}^k
    \leq
    - \sum_i \left( \tilde c_i \grad g_i^k, \v u^* \right)
    ,
    \label{eq:ineq_NS}
  \end{equation}
  where
  \begin{equation}
    F_{\rm NS}^k = \begin{cases}
      F_{\v u}^k & \textrm{for Scheme I,}\\
      F_{\v u}^k + \frac{\tau^2}{2} \norm{\frac{1}{\sqrt{\rho_0}} \grad p^k}^2 & \textrm{for Scheme II.}
    \end{cases}
    \label{eq:def_F_NS}
  \end{equation}
\end{property}
Here, the discrete kinetic energy is defined by
\begin{equation}
  F_{\v u}^k = \intV \frac{1}{2} \rho^k |\v u^k|^2 \dV .
\end{equation}
\begin{proof}
  We will first show that \cref{eq:ineq_NS} holds for Scheme I, and subsequently that it holds for Scheme II.

\paragraph{Scheme I}
  First, note that \cref{eq:single_NS1} can be written as
  \begin{multline}
    \left( \rho^{k-1} \frac{\v u^{k} - \v u^{*}}{\tau}, \v v \right)
    + \left( (\v m^{k-1} \cdot \grad ) \v u^{k} , \v v \right)
    + \int_{\partial\Omega} \lambda^k \v u^k \cdot \v v \, \diff \Gamma
    + \left( 2 \mu^k \v D \v u^{k}, \v D \v v \right)
    - \left(p^k, \div \v v \right) \\
    + \frac{1}{2} \left(  \v u^k \dpdt \rho^k, \v v \right) + \frac{1}{2} \left( \v u^k \div \v m^{k-1} , \v v \right)
    = 0 .
    \label{eq:single_NS1-2}
  \end{multline}
  Testing this with $\v v = \v u^k$ yields:
  \begin{equation}
    \frac{1}{2\tau} \norm{\sqrt{\rho^{k}} \v u^k}^2 
    - \frac{1}{2\tau} \norm{\sqrt{\rho^{k-1}} \v u^*}^2
    =
    - \norm{ \sqrt{\lambda^k} \v u^k }^2_{\partial\Omega}
    - \norm{ \sqrt{2 \mu^k} \v D \v u^{k} }^2 - \frac{1}{2\tau} \norm{\sqrt{\rho^{k-1}} ( \v u^k - \v u^*)}^2,
    \label{eq:single_NS1-3}
  \end{equation}
  since
  \begin{equation}
    \left( (\v m^{k-1} \cdot \grad ) \v u^k, \v u^k \right) + \frac{1}{2} \left( \v u^k \div \v m^{k-1}, \v u^k \right) = 0 .
  \end{equation}
  By considering \cref{eq:single_ustar}, and taking the inner product of
  it with $\rho^{k-1} \v u^*$, we obtain
  \begin{equation}
    \frac{1}{2\tau} \norm{\sqrt{\rho^{k-1}} \v u^*}^2 
    - \frac{1}{2\tau} \norm{\sqrt{\rho^{k-1}} \v u^{k-1}}^2 =
    - \sum_i \left( \tilde c_i \grad g_i^k, \v u^* \right) - \frac{1}{2\tau} \norm{\sqrt{\rho^{k-1}} (\v u^* - \v u^{k-1})}^2 .
    \label{eq:single_ustar-2}
  \end{equation}
  Summing \cref{eq:single_NS1-3,eq:single_ustar-2} yields
  \begin{multline}
    \dpdt F_{\v u}^k 
    =
 - \norm{ \sqrt{\lambda^k} \v u^k }^2_{\partial\Omega}
    - \norm{ \sqrt{2 \mu^k} \v D \v u^{k} }^2 
    - \frac{1}{2\tau} \norm{\sqrt{\rho^{k-1}} (\v u^k - \v u^*)}^2 \\
    - \sum_i \left( \tilde c_i \grad g_i^k, \v u^* \right) 
    - \frac{1}{2\tau} \norm{\sqrt{\rho^{k-1}} (\v u^* - \v u^{k-1})}^2 .
  \end{multline}
  Using \cref{eq:def_F_NS}, this yields \cref{eq:ineq_NS}.
\paragraph{Scheme II}
  The analysis for this scheme follows the same lines as in the above and closely resembles the procedure by \citet{shen2015}.

  Testing \cref{eq:split_u_1} with $\tilde{\v u}^k$ and using the definition of $\v u^*$ yields
  \begin{equation}
    \frac{1}{2\tau} \norm{\sqrt{\rho^{k}} \tilde{\v u}^k}^2 - \frac{1}{2\tau} \norm{ \sqrt{\rho^{k-1}} \v u^*}^2 + \frac{1}{2\tau} \norm{ \sqrt{\rho^{k-1}} (\tilde{\v u}^k-\v u^*)}^2
    + \norm{ \sqrt{\lambda^k} \tilde{\v u}^k }^2_{\partial\Omega}
    + \norm{ 2 \mu^k \v D \tilde{\v u}^k}^2
    = \left(p^{k-1}, \div \tilde{\v u}^k \right).
    \label{eq:uknorm1}
  \end{equation}
  Testing \cref{eq:split_p} with $\tau p^{k}$ yields
  \begin{align}
    \left( \div \tilde{\v u}^k, p^{k} \right) &= - \tau \left( \frac{1}{\rho_0} \grad (p^k - p^{k-1}) , \grad p^{k} \right) \\
                                                &= - \frac{\tau}{2} \norm{\frac{1}{\sqrt{\rho_0}}\grad p^k}^2 + \frac{\tau}{2} \norm{\frac{1}{\sqrt{\rho_0}} \grad p^{k-1}}^2 
                                                  - \frac{\tau}{2} \norm{ \frac{1}{\sqrt{\rho_0}} \grad (p^k - p^{k-1})}^2.
                                                  \label{eq:pknorm}
  \end{align}
  Testing \cref{eq:split_u_2} with $\tilde{\v u}^k$, yields:
  \begin{equation}
    \frac{1}{2\tau} \norm{\sqrt{\rho^k} \v u^k}^2 - \frac{1}{2\tau} \norm{\sqrt{\rho^k} \tilde{\v u}^k}^2 - \frac{1}{2\tau} \norm{\sqrt{\rho^k} (\v u^k - \tilde{\v u}^k) }^2 = \left( p^k - p^{k-1} , \div \tilde{\v u}^k \right) .
    \label{eq:uknorm2}
  \end{equation}
  We also have that, from Eq.\ \eqref{eq:split_u_2_scheme},
  \begin{equation}
    \norm{ \sqrt{\rho^k} (\v u^k - \tilde{\v u}^k) }^2 = \norm{ \frac{1}{\sqrt{\rho^k}} \grad ( p^k - p^{k-1})}^2 \tau^2.
    \label{eq:ukdiffnorm}
  \end{equation}
  Combination of \cref{eq:uknorm1,eq:pknorm,eq:uknorm2,eq:single_ustar-2,eq:ukdiffnorm} gives
  \begin{multline}
    \frac{1}{2\tau} \norm{\sqrt{\rho^k} \v u^k}^2 
    - \frac{1}{2\tau} \norm{\sqrt{\rho^{k-1}} \v u^{k-1}}^2 
    + \frac{\tau}{2} \norm{\frac{1}{\sqrt{\rho_0}} \grad p^k}^2 - \frac{\tau}{2} \norm{\frac{1}{\sqrt{\rho_0}} \grad p^{k-1}}^2 \\
    =
    - \frac{\tau}{2} \intV \left(\frac{1}{\rho_0} - \frac{1}{\rho^k}\right)  |\grad (p^k - p^{k-1})|^2 \, \dV
    - \sum_i \left( \tilde c_i \grad g_i^k, \v u^* \right) - \frac{1}{2\tau} \norm{\sqrt{\rho^{k-1}} (\v u^* - \v u^{k-1})}^2  \\
    - \frac{1}{2\tau} \norm{ \sqrt{\rho^{k-1}} (\tilde{\v u}^k-\v u^*)}^2
    - \norm{ \sqrt{\lambda^k} \tilde{\v u}^k }^2_{\partial\Omega}
    - \norm{ 2 \mu^k \v D \tilde{\v u}^k}^2
    .
    \label{eq:split_ekin}
  \end{multline}
  The first term on the right hand side is positive, since $\rho_0 \leq \rho^k$.
  Now, Eq.~\eqref{eq:ineq_NS} follows trivially by noting the definition \eqref{eq:def_F_NS}.
This concludes the proof.
\end{proof}
\begin{remark}
  Compared to Scheme I, the free energy in Scheme II has an extra $O(\tau^2)$ term related to pressure variations, cf.~\cref{eq:def_F_NS}.
  This is related to the weak numerical compressibility introduced by the splitting approach.
\end{remark}
\subsection{Free energy evolution for the combined steps}
\begin{property}
  \label{lemma:F}
  For the schemes presented above, the following free energy inequality holds:
  \begin{equation}
    \dpdt F^k \leq 
    - \frac{1}{\tau} \sum_i \Delta F_{c_i}^k
    + \sum_i \left( \tilde R_i , g_i^k \right)
    ,
    \label{eq:ineq_F}
  \end{equation}
  where the discrete total free energy is given by
  \begin{equation}
    F^k = F_{\rm NS}^k + F_{\rm EC}^k.
  \end{equation}
\end{property}
\begin{proof}
  This follows directly by summing \cref{eq:ineq_EC,eq:ineq_NS}.
\end{proof}


  We will now consider approximations $\tilde \alpha '(c)$ of the derivative $\alpha'(c)$ of the chemical energy $\alpha (c)$ in order to satisfy the condition $\Delta F^k_{c_i} \geq 0$, which will lead to a discrete energy stability, that is,
\begin{equation}
  \dpdt {F^k} \leq 0,
  \label{eq:energy_stable_cond}
\end{equation}
given also that the reaction terms $\tilde R_i$ are properly approximated.
The latter will be considered in Sec.\ \ref{sec:approximating_reaction}.

\subsection{Approximating the chemical energy}
\label{sec:alpha}
In the previous section, several quantities were undefined.
We now consider various numerical approximations of the chemical energy derivative $\tilde{\alpha}'$.

\paragraph{Nonlinear discretizations}
\begin{enumerate}
\item [\textbf{NL1}]
The first option is to use the non-linear approximation
\begin{equation}
  \tilde{\alpha'}(c_i^k, c_i^{k-1}) = \frac{\alpha(c_i^k) - \alpha(c_i^{k-1})}{c_i^k - c_i^{k-1}},
  \label{eq:NL1_alphaprime}
\end{equation}
which yields $\Delta F_{c_i}^k = 0$. This gives the least possible dissipation, while still leading to the correct inequality.
Note that this only holds true when the integrals are computed exactly.
On the downside, the expression \eqref{eq:NL1_alphaprime} is ill-defined when $| c^k_i - c_i^{k-1} | \ll 1$, and in order not to focus on this issue we will not consider implementations of this approximation in the present paper.
\item [\textbf{NL2}]
A second option is to use the non-linear (unless $\alpha'(c) \sim c$) approximation
\begin{equation}
  \tilde{\alpha'}(c_i^k, c_i^{k-1}) = \alpha'(c_i^{k}) .
  \label{eq:NL2_alphaprime}
\end{equation}
Taylor expansion around $c_i^k$ and the mean value theorem gives
\begin{equation}
  F_{c_i}^k - F_{c_i}^{k-1} = \int_\Omega \left[ \alpha'(c_i^{k}) (c_i^k-c_i^{k-1}) - \frac{\alpha''(\xi^k)}{2} (c_i^k-c_i^{k-1})^2 \right] \, \diff \Omega
\end{equation}
where $\xi^k \in [\min(c_i^{k-1}, c_i^{k}), \max(c_i^{k-1}, c_i^{k})]$.
This gives
\begin{align}
  \Delta F_{c_i}^k &= \tau \left( \tilde{\alpha'}(c_i^k, c_i^{k-1}), \dpdt c_i^k \right) - F_{c_i}^k + F_{c_i}^{k-1} \\
    &= \int_\Omega \frac{1}{2}\alpha''(\xi^k) (c_i^k-c_i^{k-1})^2 \, \diff \Omega .
\end{align}
Typically, $\alpha''(c) > 0$, such as for a weak solution, where $\alpha(c) = c(\log c - 1)$.
The latter leads to the common Nernst--Planck equation for the ion transport.
For such a system, where $\alpha''(c) \geq 0$ everywhere, the inequality is satisfied.
Note that if $\alpha''(c) < 0$ anywhere, a locally higher ion concentration would be favoured energetically, and effectively we could then have a negative mobility (which is mathematically ill-posed).
\end{enumerate}

\paragraph{Linear discretizations}
\begin{enumerate}
\item [\textbf{L1}]
Another option is to use the linear approximation
\begin{equation}
  \tilde{\alpha'}(c_i^k, c_i^{k-1}) = \alpha'(c_i^{k-1}) + \gamma \alpha''(c_i^{k-1})(c_i^k-c_i^{k-1}).
\end{equation}
Taylor expansion around $c_i^{k-1}$ and the mean value theorem gives
\begin{equation}
  F_{c_i}^k - F_{c_i}^{k-1} = \int_\Omega \left[ \alpha'(c_i^{k-1}) (c_i^k-c_i^{k-1}) + \frac{\alpha''(c^{k-1})}{2} (c_i^k-c_i^{k-1})^2 + \frac{\alpha'''(\xi^k)}{3!} (c_i^k-c_i^{k-1})^3 \right] \, \diff \Omega,
\end{equation}
where $\xi^k \in [\min(c_i^{k-1}, c_i^{k}), \max(c_i^{k-1}, c_i^{k})]$.
This gives
\begin{align}
  \Delta F_{c_i}^k &= \tau \left( \tilde{\alpha'}(c_i^k, c_i^{k-1}), \dpdt c_i^k \right) - F_{c_i}^k + F_{c_i}^{k-1} \\
  &= \int_\Omega \left[ \left( \gamma - \frac{1}{2}\right) \alpha''(c_i^{k-1}) - \frac{\alpha'''(\xi^k)}{3!} (c_i^k-c_i^{k-1}) \right] (c_i^k-c_i^{k-1})^2 \, \diff \Omega .
\end{align}
If $\gamma > 1/2$ the first term will be positive.
For sufficiently small $\tau$, it will dominate over the second term.
However, we have \emph{in general} no control over neither sign nor magnitude of the second term.
\item [\textbf{L2}]
To circumvent the latter problem, we may introduce a regularisation of $\alpha(c)$, denoted by $\bar\alpha(c)$.
Assuming $\alpha''(c)$ is always positive and monotonously non-increasing, we define
\begin{equation}
  \bar{\alpha}''(c) = \alpha''(\max(c, c_\delta)),
  \label{eq:regularise}
\end{equation}
where $c_\delta$ is a small cut-off concentration.
Hence $0 \leq \bar\alpha''(c) \leq \bar\alpha''(c_\delta)$.
We use the linear numerical approximation
\begin{equation}
  \tilde{\alpha}' = \bar\alpha'(c_i^{k-1}) + \left[ \gamma \bar\alpha''(c_i^{k-1}) + \frac{1}{2} \bar \alpha''(c_0)\right] (c_i^k - c_i^{k-1}),
\end{equation}
where the second term inside the brackets is a stabilizing term of order $\tau$, similar to what was used by \citet{shen2015} for the case of two-phase flow.
We expand around $c_i^{k-1}$:
\begin{equation}
  F_{c_i}^k - F_{c_i}^{k-1} = \int_\Omega \left[ \bar\alpha'(c_i^{k-1}) (c_i^k-c_i^{k-1}) + \frac{\bar\alpha''(\xi^k)}{2} (c_i^k-c_i^{k-1})^2 \right] \, \diff \Omega.
\end{equation}
This gives
\begin{align}
  \Delta F_{c_i}^k &= \int_\Omega \left[ \gamma \bar \alpha''(c_i^{k-1}) + \frac{1}{2} \left( \bar\alpha''(c_0) - \bar\alpha''(\xi^k) \right) \right] (c_i^k-c_i^{k-1})^2 \, \diff \Omega \\
    & \geq \gamma \int_\Omega \bar \alpha''(c_i^{k-1}) (c_i^k-c_i^{k-1})^2 \, \diff \Omega \geq 0,
\end{align}
where we have used that $\bar\alpha''(c_0) - \bar\alpha''(\xi^k) \geq 0$, that $\gamma \geq 0$, and that $\bar\alpha''(c_0)$.
Hence, we have derived a linear and energy stable scheme, which approximates the equations of electrohydrodynamics, given some rather general assumptions on, and a regularisation of, $\alpha(c)$.
A similar regularisation was considered recently by \citet{metzger2018}.
\end{enumerate}

In order to ensure that the whole electrochemical step is linear, it is necessary to model $\tilde K_i$ and $\tilde c_i$ to depend on the previous time step.
To this end, we will set
\begin{equation}
  \tilde K_i = \tilde K_i (c_i^{k-1}), \quad \textrm{and} \quad \tilde c_i = c_i^{k-1}.
\end{equation}
We have now considered general numerical schemes for electrohydrodynamics, and it is now necessary to give a brief summary and come with some concrete expressions.

\begin{remark}
  The regularisation defined in \cref{eq:regularise} can be applied also to the non-linear schemes to ensure that the energy is defined even if concentrations are numerically slightly negative, which might occur in simulations of highly depleted solutions, e.g.\ simulations of electrokinetic instabilities.
\end{remark}


\subsection{Approximating the reaction term}
\label{sec:approximating_reaction}
It is in place to approximate the discrete reaction term $\tilde R_i$ which enters in \eqref{eq:ineq_F}.
This term was modeled in the continuous model in \eqref{eq:reaction_R_text} and discussed in \ref{sec:modelling_reaction_terms}.
Using \eqref{eq:reaction_R_text}, we can write the discrete version as
\begin{equation}
  \tilde R_i = - \sum_m \tilde{\mathcal C}_m \sum_j \nu_{m,i} \nu_{m,j} g_j^k.
  \label{eq:R_i_discrete}
\end{equation}
Here, the reaction functions $\tilde{\mathcal C}_m$ can be modelled as $\tilde{\mathcal C}_m = \mathcal C_m^k$, i.e.\ using values from the current step, for a non-linear scheme, or as $\tilde{\mathcal C}_m = \mathcal C_m^{k-1}$, i.e., using values from the previous step, for a linear scheme.
In either case, we have that
\begin{equation}
   \sum_i \left( \tilde R_i , g_i^k \right) = -  \sum_m \norm{ \sqrt{\tilde{\mathcal C}_m} \sum_i \nu_{m,i} g_i^k }^2 \leq 0,
   \label{eq:R_i_discrete_diss}
\end{equation}
where the last equality holds given that $\tilde{\mathcal C}_m \geq 0$.
For the remainder of this article, we shall for concreteness assume the explicit treatment $\tilde{\mathcal C}_m = \mathcal C_m^{k-1}$.

\subsection{Tentative summary}
It is now appropriate to briefly summarize the major results so far.
\begin{property}
  Any decoupled scheme consisting of the combination of Scheme I or Scheme II (for the hydrodynamics), the chemical discretizations NL1, NL2 or L2, and the reaction term formulation \eqref{eq:R_i_discrete}, is energy stable.
\end{property}
\begin{proof}
  This follows from Property \ref{lemma:F} and the results for $\Delta F_{c_i}^k$ in the definitions of the discretizations NL1, NL2, L2 above, along with the result \eqref{eq:R_i_discrete_diss} for the source term.
\end{proof}
\begin{remark}
  Because of the mentioned problem with the chemical discretization L1, \emph{this approximation is not generally energy stable.}
  The discretization L1 can only be energy stable provided that $\alpha'''(c) = 0$.
\end{remark}
\begin{remark}
  When taking into account the lowest-order dissipative terms in the full expression for 
  the free energy inequality (cf.\ \cref{eq:ineq_F}), we obtain
  \begin{equation}
    \dpdt F^k \leq
    - \norm{ \sqrt{\lambda^k} \v u^k }^2_{\partial\Omega}
    - \norm{\sqrt{2 \mu^k} \v D \v u^{k} }^2
    - \sum_i \norm{ \sqrt{ \tilde K_i} \grad g_i^k }^2 ,
  \end{equation}
  which bears striking similarity with its continuous counterpart, \cref{eq:ineq_F_cont}.
  In particular, it can be verified that the terms that differ between $\dpdt F^k$ and $\pdt F$ are of order $O(\tau)$.
\end{remark}

\subsection{Concretization and specification}
The analysis thus far has considered quite general forms of the chemical energy $\alpha$, that we have presented energy-stable approximations of, the mobility $\tilde K_i$, and the chemical concentration $\tilde c_i$.
To be more specific, we therefore consider concrete forms of the undefined approximations that will be discretized and tested numerically.

\subsubsection{Chemical energy function, mobility and permittivity assumptions}
\label{sec:chemicalenergyfunction}
\begin{figure}[htb]
  \centering
  \includegraphics[width=0.9\columnwidth]{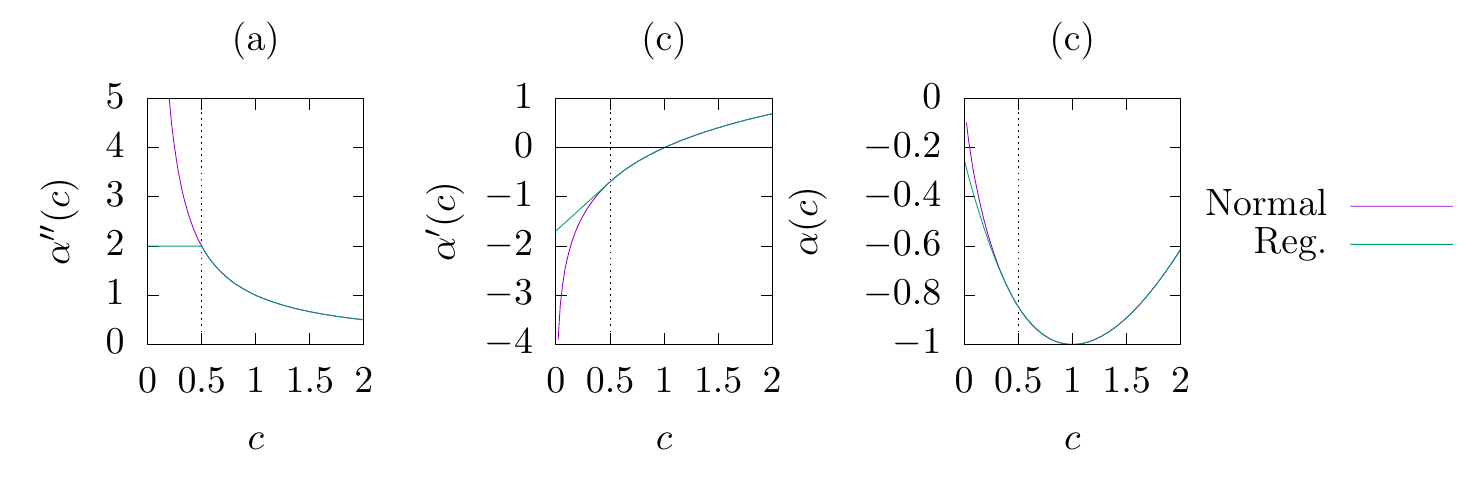}
  \caption{\label{fig:regularised}
    Regularisation of the chemical energy function $\alpha(c) = c (\ln c -1)$, with the artificially high cutoff concentration $c_\delta = 0.5$ for visual clarity.
    The cutoff concentration is indicated by a dotted vertical line.
  }
\end{figure}

We consider the Nernst--Planck equation for solute transport.
For the continuous equations, this imparts the following:
\begin{equation}
  \alpha(c_i) = c_i (\ln c_i - 1), \quad \textrm{and} \quad
  K_i(c_i) = D_i c_i,
  \label{eq:np_def_cont}
\end{equation}
where $D_i$ is the diffusion coefficient of ion species $i$.
This corresponds to dilute ionic solutions.
Since $\alpha'(c) = \ln c$ is undefined when $c \to 0$, we can regularise $\alpha$ below a small cut-off $c_\delta$, as outlined above.
Then, in the next time step, we assign $c_i^{k-1} \leftarrow \max(c_i^{k-1}, c_\delta)$.
An examplary regularisation of the functional form $\alpha(c) = c (\ln c - 1)$ is shown in Fig.\ \ref{fig:regularised}.
The regularised functional forms are:
\begin{align}
  \bar\alpha''(c) &= \frac{1}{\max(c, c_\delta)},\\
  \bar\alpha'(c) &= \begin{cases}
    \ln c & \textrm{for} \quad c > c_\delta, \\
    \ln c_\delta + \frac{c}{c_\delta} - 1 & \textrm{for} \quad c \leq c_\delta,
  \end{cases} \\
  \bar\alpha(c) &= \begin{cases}
    c (\ln c - 1) & \textrm{for} \quad c > c_\delta, \\
    c (\ln c_\delta -1)+ \frac{c^2 - c_\delta^2}{2 c_\delta} & \textrm{for} \quad c \leq c_\delta. 
    \end{cases}
\end{align}
The same regularisation was assumed by \citet{metzger2018}.

Further, we will for simplicity assume in our simulations that the permittivity does not depend on the concentrations.
Nevertheless, the schemes themselves support energy stability also in this case.

\subsubsection{Schemes used in simulations}
We define now the different schemes that will be used in simulations, and the associated approximations to \eqref{eq:np_def_cont} that will be used.
In general, the approximations should be chosen to impart soluble equation systems, i.e., for which the finite element method yields spatial convergence.


We will in this work focus on the following discretizations:

\begin{itemize}
\item[\textbf{NL2}] 
Since the discretization NL2 is non-linear, it is necessary to use e.g.~a Newton solver, where the matrices will be reassembled at each iteration, to solve this step.
A weak coupling between the Nernst--Planck and Poisson equations can be obtained by 
\begin{equation}
  \tilde K_i = D_i c_i^{k-1} \quad \textrm{and} \quad
  \tilde c_i = c_i^{k-1}.
  \label{eq:set_NL2-B}
\end{equation}


\item[\textbf{L2}]
The linear discretization in L2 imparts the following:
\begin{equation}
  \tilde K_i = D_i \max(c_i^{k-1}, c_\delta) \quad \textrm{and} \quad
  \tilde c_i = c_i^{k-1}.
\end{equation}
Without further ado, we might set $\gamma = 0$ to minimize the dissipation in this scheme.
\end{itemize}  
\begin{remark}
  A stronger coupling between the Nernst--Planck and Poisson equations in the non-linear scheme NL2, could be obtained by letting $\tilde K_i = D_i c_i^k$ and $\tilde c_i = c_i^k$.
  In general, we cannot control the sign of $\tilde K_i$ here, since we solve for $c_k$.
  Hence, if $c_k$ becomes (numerically) negative, we are not guaranteed to dissipate energy (but then the energy is not defined either).
  This issue could possibly be mitigated by a regularisation.
\end{remark}

\section{Numerical simulations}
\label{sec:results}
We have in the previous section shown how various discretization schemes satisfy a free energy inequalitity, which is also present in the models they are meant to approximate.
In this section we proceed to show and compare the effectiveness of these schemes.
The schemes have been implemented and simulations are carried out within the \emph{Bernaise} framework, developed by the authors \cite{linga2018b}.
Bernaise is a flexible simulation environment for two-phase electrohydrodynamic flow \cite{linga2018controlling}, which is built on top of the Dolfin \cite{logg2010} interface to Python within the finite element framework Fenics \cite{logg2012}.
As Fenics, \emph{Bernaise} is open-source and the latest version can be found at the online GitHub repository \cite{bernaise2018}.
Since single-phase flow is a special case of two-phase flow, \emph{Bernaise} works equally well for single-phase flow, which we consider in this paper.
For all simulations we use triangular meshes and piecewise quadratic (P2) finite elements for the velocity field, and piecewise linear (P1) elements for the remaining fields.
We use meshes that resolve the spatial problem sufficiently well for the error to be dominated by the time discretization errors.

In the following, we consider simulations of a few interesting cases.\footnote{Note the test cases considered herein are found as four separate scrips, that is, \emph{problem modules} (see \cite{linga2018b}), in the latest version of \emph{Bernaise}.
  Respectively, the problem modules are \textsf{single\_taylorgreen}, \textsf{single\_cell}, \textsf{single\_reaction}, and \textsf{single\_porous}.
  As such the results presented here are directly reproducible given a working installation of \emph{Bernaise}.
For usage of the latter, we refer to the paper describing the software \cite{linga2018b}.}
\begin{itemize}
\item First, to test the accuracy of the schemes, we consider the convergence towards an analytic solution.
\item Second, to demonstrate the energy stability of the schemes, we consider an isolated, closed system of a concentration spreading in a charged cell.
  We display the various terms in the free energy and compare the various schemes evolving in time, with varying time step $\tau$.
\item Third, we consider a reaction cell to test the reaction part of the numerical schemes.
\item Fourth, we show for a system the efficiency of the schemes to approach a steady state in an open complex geometry (porous medium) where energy is injected through a body force.
\end{itemize}

The schemes we consider are denoted by the following:
\begin{itemize}
\item \textbf{I-NL2:} Scheme I with the non-linear NL2 discretization.
\item \textbf{I-L2:} Scheme I with the linear L2 discretization.
\item \textbf{II-NL2:} Scheme II with the non-linear NL2 discretization.
\item \textbf{II-L2:} Scheme II with the linear L2 discretization.
\end{itemize}


\subsection{Accuracy test: Manufactured solution}
Now we verify the accuracy of the schemes by inspecting whether the scheme converges to a manufactured analytical solution.
Taylor--Green flow is one of a few cases for the Navier--Stokes equations where analytical solutions are available, and is therefore standard to use for validation purposes.
To this end, we consider a two-dimensional Taylor--Green flow extended to account for electrohydrodynamics.
The derivation of this manufactured solution is given in \ref{sec:derivation_manufactured}.
We consider flow of two counterions $i=\pm$, such that $z_\pm = \pm 1$, and assume constant density $\rho$, viscosity $\mu$, and permittivity $\epsilon$, and neglect gravity.

We consider the doubly periodic domain $\v x \in [0, 2\pi] \crossproduct [0, 2\pi]$, where the pressure $p$ and the  electric potential $V$ is set to zero at $\v x = (\pi/4, \pi/4)$ to fix the pressure and potential gauges, respectively.
We obtain an analytical solution augmenting \cref{eq:model_ci} with the source term $q$ on the right hand side, where
\begin{equation}
  q(x, y) = \frac{D c_0^2 C^2(t)}{2\epsilon}\left[ \cos 2x + \cos 2y + 2 \cos 2x \cos 2y \right].
\end{equation}
The analytical solution to this Taylor--Green vortex is given by:
\begin{align}
  \v u &= U(t) (\hat{\v x} \cos x \sin y - \hat{\v y} \sin x \cos y ), \\
  p &= - \frac{1}{4}\left( \rho U^2(t) + \frac{c_0^2 C^2(t)}{\epsilon} \right) (\cos 2x + \cos 2y) - \frac{c_0^2 C^2(t)}{4 \epsilon} \cos 2x \cos 2y \\
  c_\pm &= c_0 (1 \pm \cos x \cos y \, C(t)) \\
  V &= - \frac{c_0}{\epsilon} \cos x \cos y \, C(t)
\end{align}
where
\begin{align}
  U(t) &= \exp(-2\mu t/\rho), \\
  C(t) &= \chi \exp\left(-2D\left(1 + \frac{c_0}{\epsilon}\right) t \right).
\end{align}
Further, the coordinates are given by $\v x = (x, y)$ and $\hat{\v x}$ is the unit vector along $x$ and $\hat{\v y}$ is the unit vector along $y$.
A constraint ensuring that $c_\pm > 0$ is $0 \leq \chi < 1$.
The parameters used in these simulations are 
$\rho = 3$, $\mu = 2$, $D=2$, $c_0=1$, $\epsilon=2$, and $\chi=0.5$.
Further, we stop the simulation after a final time $T=0.25$, and measure the error norm respective to the analytical solution.
To rule out the error contribution from the spatial discretization, we use a fine, regular triangular mesh with diagonals from top left to bottom right, with grid size $h=2\pi/128$ (such that the discretized domain consists of $128 \crossproduct 128 \crossproduct 2$ isosceles right triangles).

\begin{figure}[htb]
  \centering
  \includegraphics[width=0.9\columnwidth]{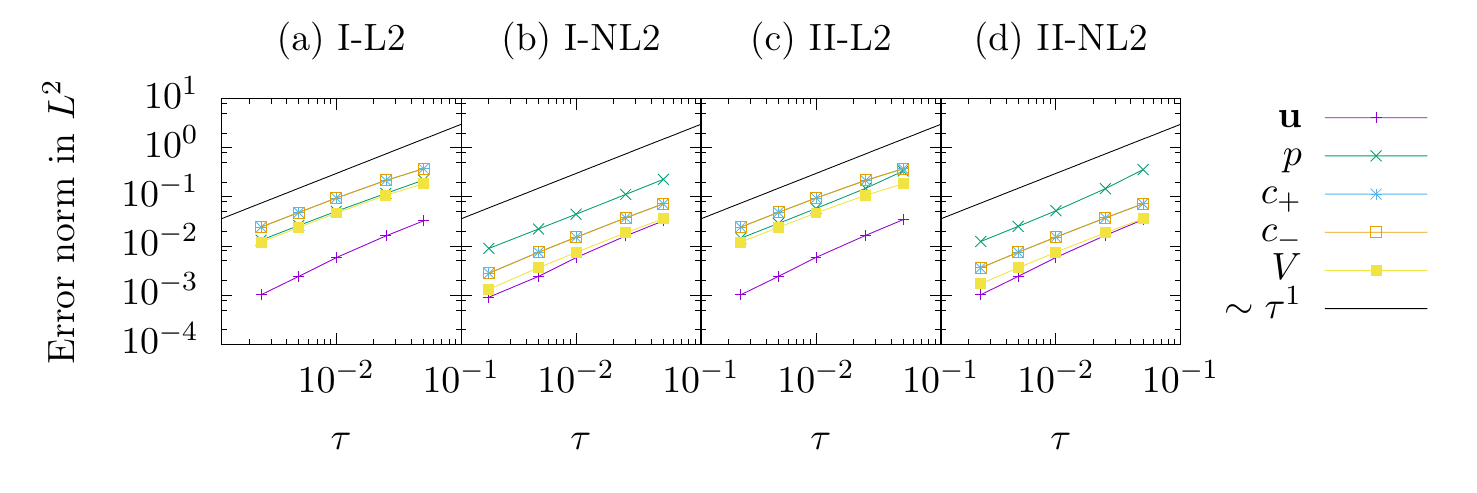}
  \caption{\label{fig:tg_convergence}
    Temporal convergence of the schemes considered in the electrohydrodynamic Taylor--Green vortex case.
    The plots (a)--(d) show the $L^2$ error norm for the various schemes for all fields compared to the reference analytical solutions as a function of time step $\tau$.
    The simulations are in good compliance with the theoretical first-order convergence prediction, indicated as a black solid slope (same in all plots).
  }
\end{figure}
In Fig.\ \ref{fig:tg_convergence}, we show convergence in the $L^2$ error norm for the four schemes considered.
Schemes I and II are virtually indistingushable.
The errors are about an order of magnitude smaller for the nonlinear NL2 scheme than for the linear L2 scheme, which is not unexpected as the NL2 provided a better approximation of the derivative of $\alpha$.
Nonetheless, all schemes seem to be reliable in that they achieve the expected $O(\tau)$ convergence.

\subsection{Stress test: Ion spreading in a charged reservoir}
\begin{figure}[htb]
  \centering
  \includegraphics[width=0.5\columnwidth]{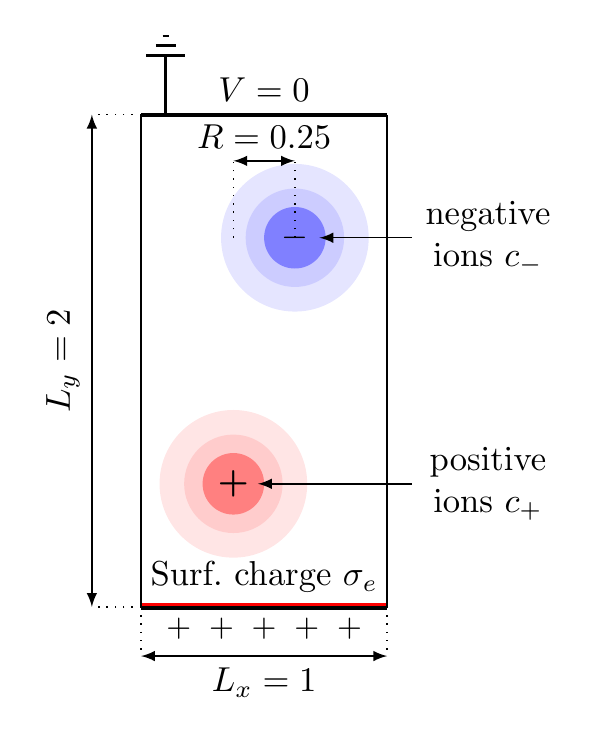}
  \caption{\label{fig:stress_test_schematic}
    Schematic set-up of the initial state in the test case of ion spreading in a charged reservoir.
  }
\end{figure}

To numerically test the energy stability of the schemes in a complex and challenging setting, we construct a system setup where the individual contributions to the free energy from inertia, chemistry and electrostatics are of comparable magnitude during the simulation.
The aim of this system is not to be physically realistic, but to reveal possible weaknesses of the schemes.
We consider a fixed domain $\Omega = [0, 1] \crossproduct [0, 2]$, which could represent a microchannel.
The geometry and initial state is sketched in Fig.~\ref{fig:stress_test_schematic}.

On the lower boundary, we assume a uniform surface charge $\sigma_e$, and the upper boundary is assumed to be grounded, i.e.~$V=0$.
The left and right boundary are assumed to be insulators.
All four walls are subject to no-slip boundary conditions on the velocity, $\v u = \v 0$.
We consider an initial state where a Gaussian concentration profile of negatively charged species is placed above and to the right, and the same profile of positively charged species is placed below and to the left of the center of the microchannel.
The initial concentration distributions are given by:
\begin{equation}
  c_\pm(x, y, t=0) = \frac{C_0}{2\pi R^2} \exp\left( - \frac{(x - L_x/2 \pm \ell_x)^2 + (y - L_y/2 \pm \ell_y)^2}{2 R^2} \right),
\end{equation}
where the parameters $L_x, L_y, \ell_x$, $\ell_y$, and $R$ are given in Table \ref{tab:ionspreading_params}.

The electrochemical interaction between the upper and lower boundaries and the two species in the bulk leads to motion due to two mechanisms.
The fluid regions with positive and negative charge are pulled (i) towards each other, and most prevalently, (ii) attracted towards opposite ends of the reservoir.
This creates a flow in the system which eventually decays due to dissipation.

The simulation parameters are listed in Table \ref{tab:ionspreading_params}.
Note that we have assumed here a linear dependency of the viscosity upon the concentrations, i.e.,
\begin{equation}
  \mu (c_\pm) = \mu_0 + \pd {\mu}{c_+} c_+ +  \pd {\mu}{c_-} c_-,
\end{equation}
where the constant coefficients $\pdinl {\mu}{c_\pm}$ are given in Table \ref{tab:ionspreading_params}.
Chosing $\pdinl {\mu}{c_\pm} \geq 0$ ensures that the viscosity is always positive.
We have also assumed a dependency of the density upon the concentration, given through the parameters $\pdinl {\rho}{c_\pm} > 0$.
As in the previous test case, we use a regular triangular mesh, now with grid size $h=1/64$ (such that the discretized domain consists of $64 \crossproduct 128 \crossproduct 2$ isosceles right triangles).

\begin{table}
\caption{\label{tab:ionspreading_params} Parameters used in the case of ion spreading in a reservoir.}
\centering
\begin{tabular}{ l c c }
  \toprule
  Parameter & Symbol & Value \\
  \midrule
  Base density & $\rho_0$ & 1.0 \\
  Base dynamic viscosity & $\mu_0$ & 0.08 \\
  Diffusivity & $D$ & 0.01 \\
  Permittivity & $\epsilon$ & 0.5 \\
  Surface charge & $\sigma_e$ & 1.0 \\
  Density per concentration & $\pdinl {\rho}{c_\pm}$ & 0.02 \\
  Dyn.\ viscosity per concentration & $\pdinl {\mu}{c_\pm}$ & 0.001 \\
  Solute mass per species & $C_0$ & 3.0 \\
  Initial spread of concentration (std.\ dev.) & $R$ & 0.25 \\
  Width of domain & $L_x$ & 1 \\
  Height of domain & $L_y$ & 2 \\
  Horizontal displacement of initial conc. & $\ell_x $ & 0.125 \\
  Vertical displacement of initial conc. & $\ell_y $ & 0.5 \\
  Total simulation time & $T$ & 10 \\
  Cut-off concentration (only L2) & $c_\delta$ & 0.1 \\
  \bottomrule
\end{tabular}
\end{table}

\begin{figure}[htb]
  \centering
  \includegraphics[width=0.8\columnwidth]{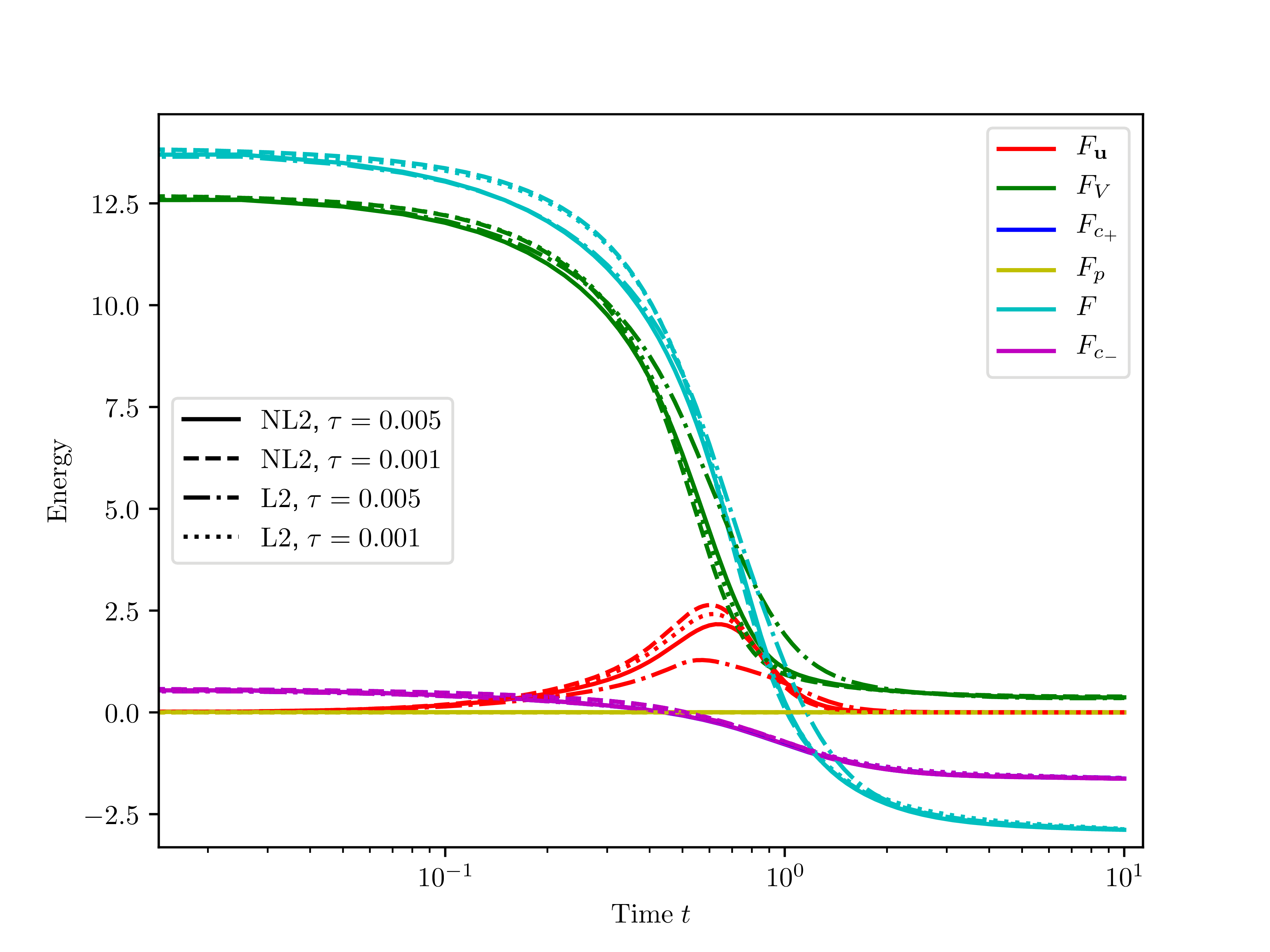}
  \caption{\label{fig:energy_cell}
    Free energy in time.
    All simulations are done using fractional step hydrodynamics, i.e., Scheme II.
    Note that, cf.\ Eq.\ \eqref{eq:def_F_NS} for Scheme II, we have defined $F_p = \tfrac{1}{2} \rho_0^{-1} \tau^2 \norm{ \grad p^k}^2$, which represents a compressive energy related to the splitting, and has a negligible contribution to the total free energy $F$ for these simulations. ($F_{c_-}$ has been omitted, as it is identical to $F_{c_+}$.)
  }
\end{figure}

We first inspect the evolution of the total free energy in time, which should unconditionally decrease for energy-stable schemes.
The various contributions to the total free energy, integrated over the domain, are shown in Fig.\ \ref{fig:energy_cell}.
Here, we have compared the two chemical discretization strategies L2 and NL2, and two time step sizes.
From the latter figure, it is evident that the schemes approach the same equilibrium state regardless of the time step size $\tau$ and discretization.
We observe that the increased dissipation due to a larger time step size results in lower fluid speed, which in turn leads to delayed equilibration.
Moreover, as expected, the linear L2 scheme is more dissipative than the NL2 scheme and requires much a smaller time step to produce a reliable kinetic energy development, cf.\ Fig.\ \ref{fig:energy_cell}.
Nonetheless, the schemes always decrease the total free energy in every time step, as expected.

\begin{figure}[htb]
  \centering
  \includegraphics[width=0.9\columnwidth]{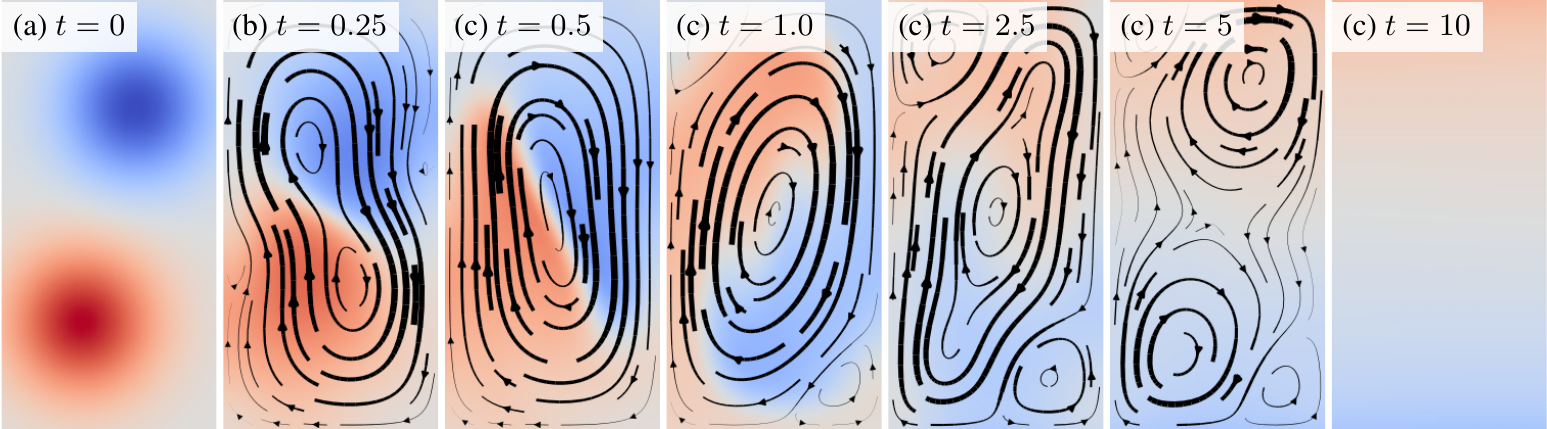}
  \caption{\label{fig:snapshots_cell}
    Snapshots in time of the ion spreading simulation case.
    The flow lines are normalized for each simulation and omitted in the first and last snapshots.
    The color indicates the net charge, red is positive, blue is negative, and gray is neutral.
    The related color scale is normalized for the entire simulation.
    For this simulation, Scheme II-NL2 with a time step $\tau=0.005$ was used.
  }
\end{figure}

It is also interesting to investigate visually how the flow behaves in this setup.
In Fig.\ \ref{fig:snapshots_cell} we show snapshots from a simulation of this system at several instances of time.

\subsection{Reaction cell}

\newcommand{\rmA}{{\textrm{A}^+}}
\newcommand{\rmB}{{\textrm{B}^-}}
\newcommand{\rmAB}{{\textrm{AB}}}
To verify the modelling and implementation of the reaction term, we now simulate a reaction cell test case.
We consider the simple reaction
\begin{equation}
  \rmA + \rmB \rightleftharpoons \rmAB.
\end{equation}
We define $c_\rmA, c_\rmB$ and $c_\rmAB$ to be the associated concentrations.
The associated stoichiometric coefficients are now $\nu_\rmA = \nu_\rmB = -1$ and $\nu_\rmAB = 1$.
We let the reference concentrations (at equilibrium) be defined by $c_\rmA^0 = c_\rmB^0 \equiv c^0 = 3$ and $c_\rmAB^0 = 1$.
We consider reaction kinetics as the example discussed in \ref{sec:modelling_reaction_terms}, i.e.,
\begin{equation}
  \mathcal{C} = \mathcal{C}_0 \frac{e^{g_\rmAB} - e^{-g_\rmA -g_\rmB}}{g_\rmAB -g_\rmA - g_\rmB},
\end{equation}
which is a generalization of the law of mass action.
Here, $\mathcal{C}_0$ is a constant coefficient.
The same reaction kinetics was considered, e.g., by \citet{campillo-funollet2012,metzger2018}.
Hence, in equilibrium, we should have
\begin{equation}
  g_\rmAB -g_\rmA - g_\rmB = 0, \quad \textrm{which gives} \quad \frac{c_\rmA \cdot c_\rmB} {c_\rmAB} = \frac{\left( c^0 \right)^2} {c_\rmAB^0} = K_{\rm sp}^{-1} = 9.
\end{equation}
We consider a domain $\Omega = [-0.5, 0.5] \crossproduct [-0.5, 0.5]$, where we start out the simulation with a Gaussian distribution of neutral species $\rmAB$ centered at $(0, 0)$ and with a standard deviation $R=0.15$:
\begin{equation}
  c_\rmAB (x, y, t=0) = \frac{c_0}{2\pi R^2} \exp\left( - \frac{x^2 + y^2}{2 R^2} \right).
\end{equation}
At the bottom boundary we apply a surface charge $\sigma_e$, and the top boundary is grounded.
At the left and right boundary we apply no-flux conditions, and all boundaries are subject to the no-slip condition $\v u = \v 0$ on the velocity field.
We take the initial average concentration of the chemical species $\rmAB$ in the domain to be $c_0 = 10$.
The other ions are set initially to a constant (negligibly) low concentration $c_\rmA = c_\rmB = 10^{-4}$.
Hence, in the absence of an applied electric field, the uniform equilibrium concentrations should be $c_\rmA = c_\rmB = 6$ and $c_\rmAB = 4$.

The equilibrium state with an applied electric field is also possible to find quasi-analytically.
The solution will thus only depend on the vertical coordinate $y$.
We consider a domain $y \in [-\ell, \ell]$.
At equilibrium, the electrochemical potentials must be constant:
\begin{equation}
  g_i = \ln \left( \frac{c_i (y)}{c_i^0} \right) + z_i V(y) = \textrm{const.}
\end{equation}
Without loss of generality, we take the electrostatic potential $V(y)$ to be antisymmetric about $y=0$ (and thus omit the grounded boundary condition at the top).
Thus, $V(0) = 0$.
Further, due to symmetry, the concentrations $c_\rmA(0) = c_\rmB(0) \equiv \bar c$ (const.) here.
Therefore, the constant $g_i = \ln \left({\bar c}/{c_i^0} \right)$ for $i \in \{\rmA, \rmB\}$, and
\begin{equation}
  c_i(y) = \bar c e^{-z_i V(y)} \quad \textrm{for} \quad i \in \{\rmA, \rmB\}.
\end{equation}
The neutral concentration will be uniform, i.e., $c_\rmAB = K_{\rm sp} \bar c^2$.
This gives, in the Poisson equation,
\begin{equation}
  \epsilon \dd {V} y = -c_\rmA + c_\rmB = 2 \bar c \sinh(V),
  \label{eq:npb}
\end{equation}
where we still need to determine the value of the unknown constant $\bar c$.

The average number of ions must be conserved.
We started out with an average concentration $c_0$ of only $\rmAB$ which contains both $\rmA$ and $\rmB$.
Conservation of both ions can, e.g., be written as:
\begin{equation}
  c_\rmAB + \frac{1}{2\ell} \int_{-\ell}^{\ell} \frac{c_\rmA + c_\rmB}{2} \diff y = c_0,
\end{equation}
since we have already assumed that that the total number of ions of $\rmA$ and $\rmB$ is equal.
Inserting for $c_\rmAB$ and $c_\rmA, c_\rmB$, we get
\begin{equation}
  K_{\rm sp} \bar c^2 + \bar c \int_{-\ell}^\ell \cosh(V) \, \diff y = c_0. 
  \label{eq:integro}
\end{equation}
The charged boundary condition can be written as
\begin{equation}
  \d V y = - \frac{\sigma_e}{\epsilon}
  \label{eq:npb_neumann}
\end{equation}
at both the upper and the lower boundary.
We thus have to solve the nonlinear Poisson--Boltzmann equation \eqref{eq:npb} with the Neumann boundary conditions \eqref{eq:npb_neumann} coupled with the integral \eqref{eq:integro}.
This can be done numerically with standard ordinary differential equation solvers.
With the chosen parameters, we obtain $F_V = 1.5516$, $F_{c_\rmA} = F_{c_\rmB} = -0.6890$ and $F_{c_n} = 0.9927$.

We choose also the dynamic parameters $D=0.01$, $\mathcal{C}_0 = 10$, $\pdinl {\rho} {c_\rmA} = \pdinl {\rho} {c_\rmB} = 0.1$, $\pdinl {\rho} {c_\rmAB} = 0.2$, $\pdinl {\mu} {c_\rmA} = \pdinl {\mu} {c_\rmB} = 0.02$, $\pdinl {\mu} {c_\rmAB} = 0.04$, a time step $\tau = 0.01$ and a total simulation time $T=10$.
For the spatial discretization, we use a uniform grid size $h = 1/128$ with isosceles right triangles whose diagonals all go from top left to bottom right. 

\begin{figure}[htb]
  \centering
  \includegraphics[width=0.7\columnwidth]{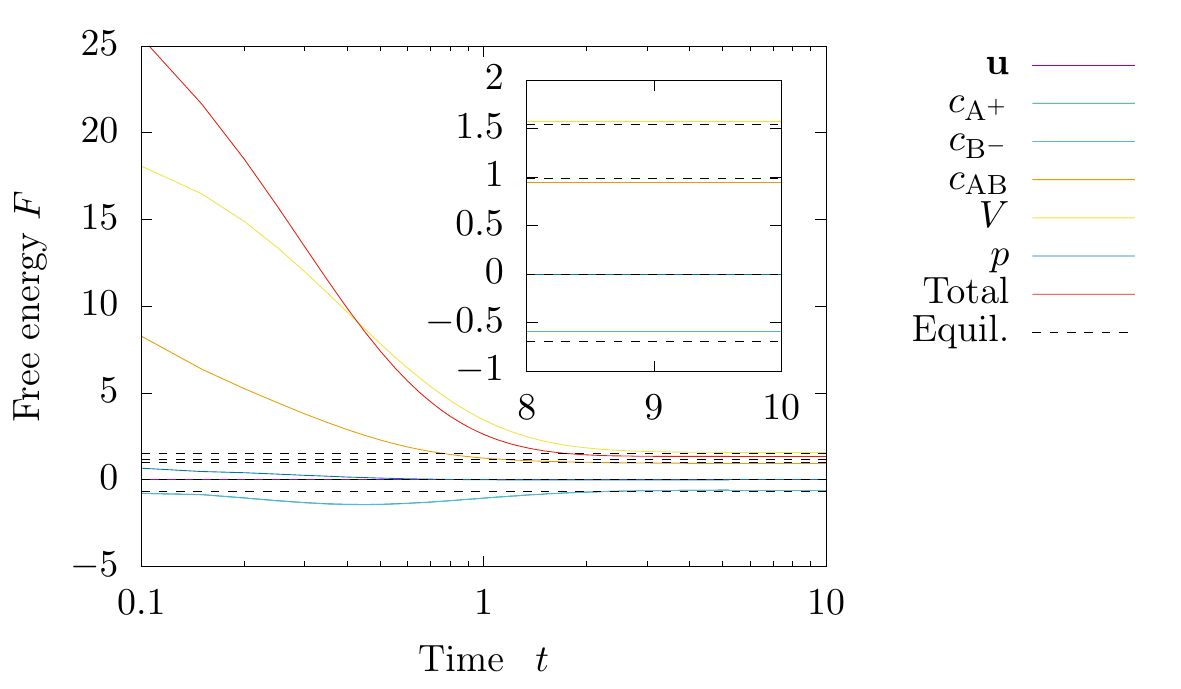}
  \caption{\label{fig:reaction_cell}
    Free energy in time for the reaction cell simulation case.
    These simulations were carried out using Scheme II-NL2.
    The inset shows a close-up of the data (except the total energy, for clarity).
  }
\end{figure}

In Fig.\ \ref{fig:reaction_cell} we demonstrate how the energy decays towards the calculated energy values for the scheme II-NL2.
As shown in the inset, the values are fairly close to the equilibrium values although we have not simulated many diffusive time scales.
Therefore the (total) chemical energy is slightly above the equilibrium values.
The other schemes yield similar results, but are omitted in the figure for visual clarity.

\subsection{Application: Electrohydrodynamic flow in a charged porous medium}
\label{sec:porous}
Finally, we test the applicability of the schemes in a case where energy is injected into the system.
The overall discrete free energy inequality will then be broken.
Energy stable schemes are nevertheless useful, since the dissipation guarantee in the bulk will still hold.
The departure from global energy dissipation will be controlled by the flux through the inlet and the outlet of the system.

We consider flow in a two-dimensional domain $\Omega = \{ (x, y) \in [-L_x/2,L_x/2] \crossproduct [-L_y/2, L_y/2]\}$, where $L_x, L_y$ are domain size along the $x, y$ directions, respectively, and $L_x > L_y$.
The domain is taken to be periodic in the $y$-direction.
Within the domain, there are $N=8$ circular obstacles with radius $R$ placed randomly within the subdomain $[-L_y/2, L_y/2]\crossproduct[-L_y, L_y/2]$, but no closer to any other obstacle than $R$.
We assume the no-slip boundary condition, $\v u = \v 0$, on the obstacles, and $p=0$ on left and right boundaries.
The flow is driven by an average pressure gradient, implemented as a uniform body force $\v f_b = f_b \hat{\v x}$.
Further, a constant concentration $c_+ = c_- = c_0$ is assumed at both inlet and outlet.
The left side is grounded, $V=0$, and on the right side we assume a no-flux condition on the electric field, $\hat{\v n} \cdot \grad V = 0$.
These boundary conditions are fairly standard in this kind of computation \cite{mansouri2005,mansouri2007,bolet2018}.

We will now compare the time-dependent solution using the schemes presented herein to the steady-state solution provided by the independently developed solver presented in a preceding paper by the authors \cite{bolet2018}.
The simulations parameters are given in Table \ref{tab:params}.
An unstructured triangular mesh, with a fine typical linear size $h \approx 0.25$ was used to minimize errors from the spatial discretization.
In particular, the mesh consists of 78280 triangles and 39898 nodes.
Based on the resulting maximum velocity $U \simeq 3 \cdot 10^{-1}$, the pore radius $R$, and the kinematic viscosity $\mu/\rho$, we can estimate the Reynolds number to be $\mathrm{Re} = \rho U R / \mu \simeq 0.02$.
Further, the Schmidt number can be estimated to $\mathrm{Sc} = \mu /(\rho D) \simeq 100$, and P\'{e}clet number $\mathrm{Pe} = U R / D = \mathrm{Re} \cdot \mathrm{Sc} \simeq 2$.
We can also estimate the Debye length in these units to be $\lambda_D = \sqrt{\epsilon/(2 c_0)} \simeq 1.5$, i.e., the dimensionless Debye length to pore size is $\lambda_D/R \simeq 0.5$.

\begin{table}[htb]
  \caption{\label{tab:params}
    Parameters used in the simulations of electrohydrodynamic flow in a charged porous medium.}
  \centering
  \begin{tabular}{l l l}
    \toprule
    Parameter & Symbol & Value \\
    \midrule
    Domain length along $x$  & $ L_x $ & $ 60 $ \\
    Domain length along $y$ (periodic direction) & $ L_y $         & $ 30 $ \\
    Number of obstacles      & $ N $           & $ 8 $ \\
    Obstacle radius          & $ R $           & $ 3.0 $ \\
    Concentration            & $ c_0$          & $ 1 $ \\
    Surface charge           & $ \sigma_e $    & $ -5 $ \\
    Density                  & $ \rho $        & $ 0.02 $ \\
    Dynamic viscosity        & $ \mu $         & $ 4.5 $ \\
    Permittivity             & $ \epsilon $    & $ 4.5 $ \\
    Diffusivity of ions      & $ D $           & $ 0.457 $ \\
    Average pressure gradient & $ f_b $           & $ 0.09 $ \\
  \bottomrule
  \end{tabular}
\end{table}

\begin{figure}[htb]
  \centering
  \sidesubfloat[]{\includegraphics[width=0.65\columnwidth]{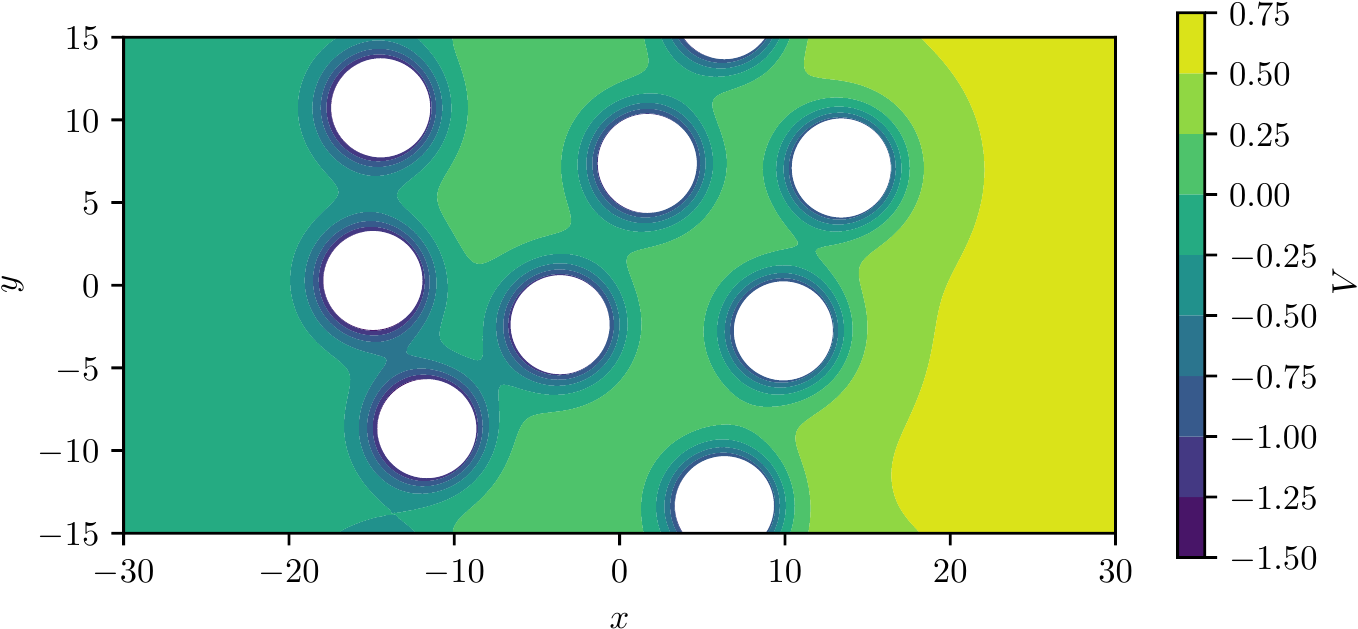}\label{fig:V}}\\
  \sidesubfloat[]{\includegraphics[width=0.64\columnwidth]{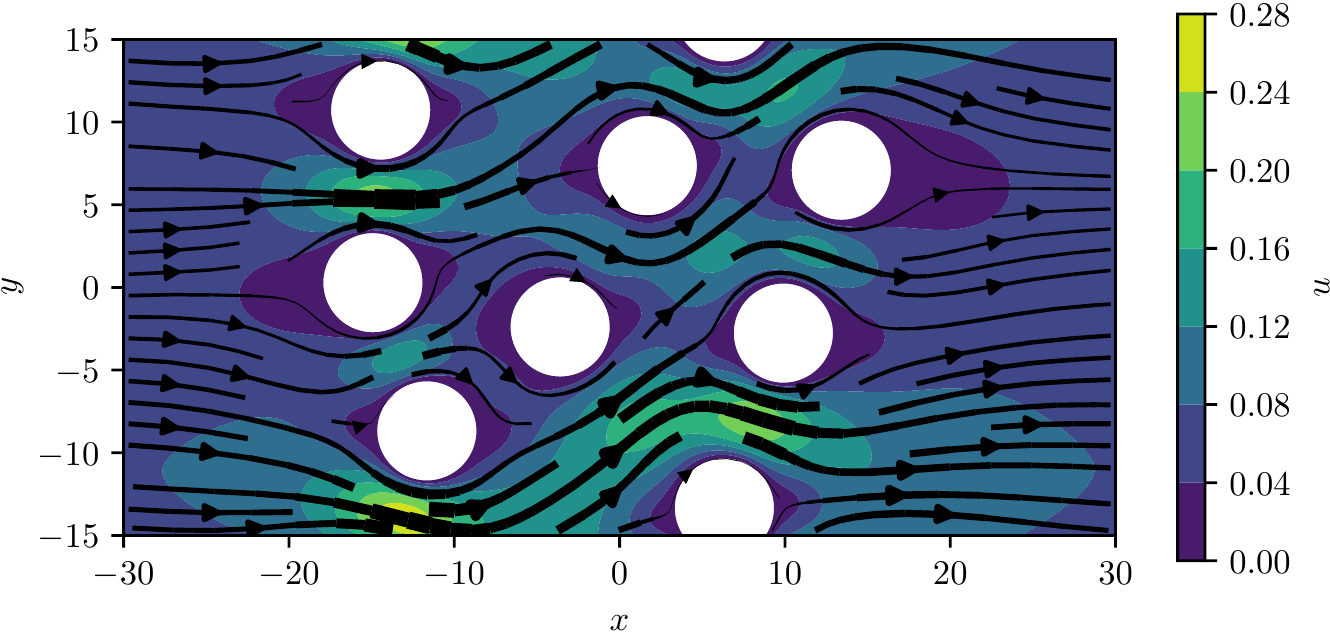}\label{fig:u}}
  \caption{%
    Steady state fields for the case of electrohydrodynamic flow in a porous medium.
    (a) Electric potential.
    (b) Velocity field.
  }
\end{figure}

\begin{figure}[htb]
  \centering
  \includegraphics[width=0.6\columnwidth]{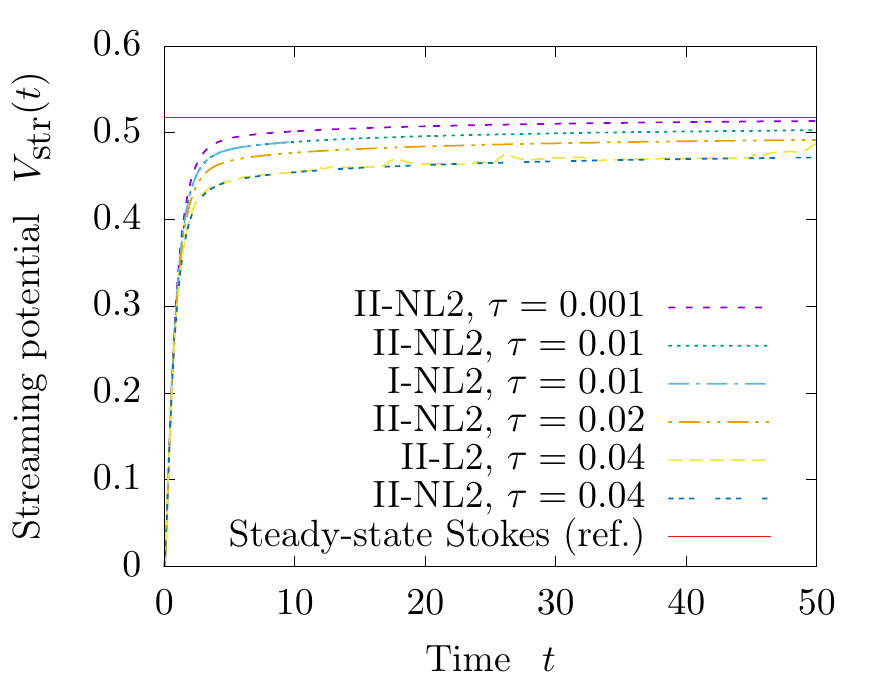}
  \caption{\label{fig:V_approach}
    We show the time-development of the streaming potential, comparing the time-dependent solution $V(t)$ to the reference steady-state solution $V_{\rm ref}$ obtained by the method presented in \cite{bolet2018}.
    The time-dependent solution relaxes exponentially to the steady-state solution.
  }
\end{figure}

The steady-state solver was run with the same settings as the time-dependent solver, only differing in the fact that the velocity field is periodic also in the $x$-direction (while the ionic system is finite in the $x$-direction), and that the inertial term is completely ignored ($\mathrm{Re}=0$).
Hence, this steady-state should represent a minimum of dissipation.
The electric potential of the steady-state solver is presented in Fig.\ \ref{fig:V} and the velocity field is shown in Fig.\ \ref{fig:u}.
In Fig.\ \ref{fig:V_approach}, we measure in time the potential at the right boundary, i.e.~the streaming potential, as a function of time, obtained with the various time-dependent schemes.
Also plotted is the reference streaming potential obtained with the steady-state solver.
The total simulation time is $T=50$.
We may define a diffusive time scale $\tau_D$ based on the Debye length, $\tau_D = \lambda_D^2 / D \simeq 5$; hence we have simulated here over about 10 of this diffusive time scale.
This time scale may be present in the fast decay seen in the initial stages in Fig.\ \ref{fig:V_approach}.
From Fig.\ \ref{fig:V_approach}, it is clear that the time step $\tau$ has a relatively strong effect on the resulting streaming potential.
In particular, the $O(\tau)$ dissipative term that will be present in the steady state, due to the presence of $\v u^*$ in the scheme, has consequences also for the streaming potential.
Hence, good agreement is only found for relatively fine time steps.
Finally, we conclude from this figure that the linear EC scheme L2 is less precise than the NL2 scheme, and hence NL2 may be required for this type of computation.
For this particular problem, there does not seem to be a pronounced difference between the coupled and the splitting scheme.

\section{Discussion and conclusion}
\label{sec:discussion}
The contribution of the work presented here is twofold.
Firstly, we have presented a general model for single-phase electrohydrodynamic flows, where the fluid properties are allowed to depend on the concentrations of ions.
Secondly, we have proposed discretization strategies for the resulting set of equations.
The proposed schemes impart decoupled computation of electrochemistry and hydrodynamics, while still satisfying the same free energy inequality as the underlying model.
In particular, as opposed to schemes that do not satisfy the latter property, this guarantees that we do not violate basic physical features of the underlying model regardless of time step size.

The results presented allow for the following discussion.
\begin{itemize}
\item The model presented in this work is fairly general, and provides a consistent way of including permittivity gradients, gravitational effects and viscosity dependence on salinity in simulations of electrohydrodynamics.
  This also imparts that the model can be used to study simplified systems, such as the effects of salinity gradients in the absence of electric fields.
  Further, the effects of non-constant density and permittivity can be included in studies of electrokinetic instabilities beyond the Boussinesq approximation (see e.g.\ \cite{karatay2016}).
\item The limitations of the model are (i) that we have assumed quasi-incompressibility (solenoidal velocity field), and (ii) that we have assumed isothermal flow.
  The first assumption is commonplace even beyond the Boussinesq approximation, see e.g., \cite{szewc2011,wu2017}.
  The second is standard in electrokinetics.
\item The assumption of a linear equation of state \eqref{eq:eos_rho_cj} combined with a solenoidal velocity field \eqref{eq:model_NS2} can in some settings be overly restrictive, particularly for less dilute solutions.
  Ideally, more general density descriptions should be considered.
  In this respect, the model by \citet{dreyer2013} is particularly relevant, and could be a starting point for further improvement.
  Instead of \cref{eq:eos_rho_cj,eq:model_NS2} they assume a constant number density to close the model, while real systems would be located somewhere between the two models.
  It should be stressed that developing stable and efficient schemes for such models would remain an important and challenging research topic.
\item Dependence on the electric field strength, in particular for the permittivity, has been ignored in the model, although studies indicate that it might be significant at high field strengths \cite{pride1991,booth1951}.
  It is in principle trivial to include this effect by letting $\epsilon$ be a function of $|\v E|^2$ (as well as $\{c_i \}$) in \eqref{eq:pd_E}.
\item The decoupling strategy is highly efficient, in the sense that it permits the use of specialized numerical routines for the resulting subproblems.
  Hence, the schemes should facilitate efficient simulations of electrohydrodynamic flows in arbitrary complex geometries.
\item In particular, the fractional-step method (Scheme II) for the hydrodynamics leads to significant speed-up compared to the coupled hydrodynamics (Scheme I).
  Combined with the linear chemical discretization L2, which is based on a regularisation and a stabilization of the chemical potential, it yields a completely linear scheme that can be solved at each time step.
\item Since the velocity field will typically have to be resolved with a higher spatial order than the pressure field (e.g., P2-P1 elements for the mixed problem) to deal with the Babuszka--Brezzi condition \cite{brenner2007}, the main computational cost may still be associated with computing the velocity field.
  In these cases, choosing a nonlinear chemical discretization (e.g., NL2) might be worthwhile, as it gives a more accurate solution while not contributing significantly to the computational runtime.
  The results shown in Sec.\ \ref{sec:porous} underpin this observation.
\item The decoupling between electrochemistry and hydrodynamics introduces a time step restriction (related to the Courant number), since the advective term in the chemical transport equation is integrated explicitly.
  Thus, fully implicit methods will possibly be more stable, allowing larger time steps, and may for certain applications be more efficient.
\item The work presented here, in particular related to the numerical schemes, builds on many known results from the literature, e.g.\ \cite{minjeaud2013,guillen-gonzalez2014,metzger2015,shen2015,metzger2018}.
  A main novelty in the present work is to combine the results on chemical potential stabilization and fractional schemes known from phase-field simulations of two-phase flows \cite{shen2015} with electrochemical gradients \cite{metzger2015,metzger2018}.
  Further, these methods have been adapted to the case where fluid properties depend on concentrations rather than an order parameter (phase) field.
\item Rigorously proving existence of solutions and convergence of the proposed numerical schemes is a challenge that has not been undertaken in the present work.
  Progress here could be made along the lines of related work, see e.g., Ref.\ \cite{shen2015,metzger2018}.
\end{itemize}
In future work, the model and scheme should be generalized to multiphase systems.
In particular, this would impart a combination of the present work and the model by \citet{campillo-funollet2012}.
To simulate solid-liquid interaction, the geometry could be described by a phase field which could evolve due to chemical reactions at the interface, i.e., the function $\mathcal{C}$ could be nonzero only here.
Then phase transformations from solute to could occur only at the phase field interface and proportionally (or another functional dependence) to the concentration of a given species.
This could provide a refinement to other studies \cite{xu2008,hawkins2014}.

A more challenging, but highly physically relevant, extension of the model would be to extend it to encompass both non-isothermal flow and non-solenoidal velocity fields.
This would require a derivation taking into account entropy production rather than free energy dissipation.
Non-solenoidal velocity fields would also require more sophisticated numerical schemes for reliable and efficient simulation.


\appendix


\section{Modelling the reaction terms}
\label{sec:modelling_reaction_terms}
Here we consider the modelling of the reaction terms $R_i$.
The dissipation related to the reaction is given by (cf.\ \eqref{eq:dFdt_general})
\begin{equation}
  \sum_i \intV g_i R_i \, \dV.
  \label{eq:diss_react}
\end{equation}
We consider a set of $M$ possible reactions including all $N$ chemical species, where we can write the reactions in the following way:
\begin{gather*}
  \nu_{1,1} \chi_1 + \ldots + \nu_{1,N} \chi_N \rightleftharpoons 0, \\
  \vdots \\
  \nu_{M,1} \chi_1 + \ldots + \nu_{M,N} \chi_N \rightleftharpoons 0,
\end{gather*}
where $\chi_i$ symbolizes the chemical species, and $\nu_i$ is the corresponding \emph{net} stoichiometric coefficent.
The latter is such that $\nu_i > 0$ for (net) products and $\nu_i < 0$ for (net) reactants.
If the chemical species does not enter into the reaction, $\nu_i = 0$. 
More compactly, we can write
\begin{equation}
  \sum_i \nu_{m,i} \chi_i \rightleftharpoons 0, \quad \forall m \in [1, M].
  \label{eq:reaction_general}
\end{equation}
Note that due to charge conservation in a reaction, $\sum_i z_i \nu_{m,i} = 0$ and due to mass conservation in a reaction, $\sum_i \nu_{m,i} \pdinl{\rho}{c_i} = 0$, for all reactions $m$.
For each reaction $m$ we have a reaction rate $\mathcal{R}_m$.
The reaction source term that enters in the concentration equation of species $i$, can be written as
\begin{equation}
  R_i = \sum_m \nu_{m,i} \mathcal R_m.
  \label{eq:reaction_Ri}
\end{equation}
Now, what remains is to define $\mathcal R_m$ on physical grounds.
We have from statistical mechanics that in equilibrium, the reaction \eqref{eq:reaction_general} is given by
\begin{equation}
  \sum_i \nu_{m,i} g_i^0 = 0,
\end{equation}
where the superscript ``0'' indicates \emph{local} equilibrium.
This suggests that a form
\begin{equation}
  \mathcal{R}_m = - \mathcal{C}_m \cdot \sum_i \nu_{m,i} (g_i - g_i^0) = - \mathcal{C}_m \cdot \sum_i \nu_{m,i} g_i,
  \label{eq:reaction_Rm}
\end{equation}
where $\mathcal C _m \geq 0$, should drive the species towards equilibrium; in the sense that
\begin{itemize}
\item a term with $g_i > g_i^0$ should promote generation of more reactants ($\nu_{m,i} < 0$) and less products ($\nu_{m,i} > 0$),
\item a term with $g_i < g_i^0$ should push towards less reactants and more products, and
\item a term with $g_i = g_i^0$ should not contribute.
\end{itemize}
Inserting \eqref{eq:reaction_Ri} and \eqref{eq:reaction_Rm} into \eqref{eq:diss_react},
\begin{align}
  \sum_j \intV g_j R_j \, \dV &= - \sum_m \mathcal{C}_m \sum_i \sum_j \intV g_j \nu_{m,j} \nu_{m,i} g_i \, \dV \\
                              &= - \sum_m \mathcal{C}_m \intV \left( \sum_i \nu_{m,i} g_i \right)^2 \, \dV \leq 0,
\end{align}
which is clearly dissipative.

Note that in general, no assumptions were made about $\mathcal C_m$ except that it should be nonnegative.
For dilute systems described by the classical Nernst--Planck equations this is in general satisfied.
Here, $g_i = \ln c_i - \ln c_i^0 + z_i V$, and in general, we can model by statistical rate theory:
\begin{align}
  \mathcal R_m &= - k_{\textrm{b}, m} \prod_{\nu_{m,i} > 0} c_i^{\nu_{m,i}} + k_{\textrm{f}, m} \prod_{\nu_{m,i} < 0} c_i^{-\nu_{m,i}} \\
               &= \left[ - k_{\textrm{b}, m} \prod_{\nu_{m,i} > 0} (c_i^0)^{\nu_{m,i}} e^{ \sum_{\nu_{m,i} > 0} g_i \nu_{m,i} } + k_{\textrm{f}, m}  \prod_{\nu_{m,i} < 0} (c_i^0)^{-\nu_{m,i}} e^{ - \sum_{\nu_{m,i} < 0} g_i \nu_{m,i} } \right] e^{-\sum_{\nu_{m,i} > 0} z_i \nu_{m,i} V}.
\end{align}
Here, $k_{\textrm{f},m]}$ is the forward reaction rate and $k_{\textrm{b},m}$ the backward rate.
The references $c_i^0$ are defined through the equilibrium condition
\begin{align}
  0 &= - k_{\textrm{b}, m} \prod_{\nu_{m,i} > 0} (c_i^0)^{\nu_{m,i}} + k_{\textrm{f}, m}  \prod_{\nu_{m,i} < 0} (c_i^0)^{-\nu_{m,i}},
\end{align}
which relates to the solubility product $K_{\rm sp}$ through the law of mass action,
\begin{equation}
  K_{{\rm sp}, m} = \frac{k_{\textrm{f}, m}}{k_{\textrm{b}, m}} = \frac{\prod_{\nu_{m,i} > 0} (c_i^0)^{\nu_{m,i}}}{\prod_{\nu_{m,i} < 0} (c_i^0)^{-\nu_{m,i}}}.
\end{equation}
Inserting into the above,
\begin{align}
  \mathcal R_m &= - k_{\textrm{b}, m} \prod_{\nu_{m,i} > 0} (c_i^0)^{\nu_{m,i}} e^{- z_i \nu_{m,i} V}\left[ e^{\sum_{\nu_{m,i} > 0} g_i \nu_{m,i} } - e^{ - \sum_{\nu_{m,i} < 0} g_i \nu_{m,i} } \right] \\
  &= - k_{\textrm{b}, m} \prod_{\nu_{m,i} > 0} (c_i^0)^{\nu_{m,i}} e^{- z_i \nu_{m,i} V} \frac{ e^{\sum_{\nu_{m,i} > 0} g_i \nu_{m,i} } - e^{ - \sum_{\nu_{m,i} < 0} g_i \nu_{m,i} } }{\sum_i g_i \nu_{m,i}} \sum_i g_i \nu_{m,i} \\
  &= - \mathcal C_m \sum_i g_i \nu_{m,i} . 
\end{align}
Where we have identified
\begin{equation}
  \mathcal C_m = k_{\textrm{b}, m} \prod_{\nu_{m,i} > 0} (c_i^0)^{\nu_{m,i}} e^{- z_i \nu_{m,i} V} \frac{ e^{\sum_{\nu_{m,i} > 0} g_i \nu_{m,i} } - e^{ - \sum_{\nu_{m,i} < 0} g_i \nu_{m,i} } }{\sum_i g_i \nu_{m,i}}
\end{equation}
Note that for any $x_1, x_2 \in \mathbb R$,
\begin{equation}
  \zeta(x_1)-\zeta(x_2) = \zeta'(x) (x_1-x_2),
\end{equation}
for some $x\in [\min(x_1, x_2), \max(x_1, x_2)]$.
Since $[\exp(x)]' \geq 0$ for all $x$, we have that $\mathcal C_m \geq 0$.

\section{Proof of non-negative concentrations}
\label{sec:proof-non-negative}
Here we establish that the concentrations are always non-negative for the case when the mobility is modelled as $K_i = D_i c_i$.
For the case of standard Nernst--Planck transport, this result was shown by \citet{schmuck2009}.
To this end, we follow in the lines of \cite{schmuck2009} and introduce an auxiliary problem,
\begin{equation}
  \pdt c + \v u \cdot \grad c - \div \left( (c)_+ \grad g ( (c)_+, V) \right) = R^+ - R^- (c)_+,
  \label{eq:c_aux}
\end{equation}
where $(\cdot)_+ = \max(0, \cdot)$.
The contributions to the reaction terms are $R^+, R^- \geq 0$, such that the first term on the right hand side represents the creation of the species $c$ from other chemicals and the second term represents the removal of it.
We now define $C^+ = (c)_+$ and $C^- = (-c)_+$, such that we can write $c = C^+ - C^-$.
Testing \cref{eq:c_aux} with $C^-$ and integrating by parts yields
\begin{equation}
  -\frac{1}{2} \d{}{t} \norm{C^-}^2 = - \left(C^+ \grad C^-, \grad g \right) + \left(R^+, C^-\right) + \left(R^- C^+, C^-\right) = \left(R^+, C^-\right).
\end{equation}
Integrating over time $s \in [0, t]$ yields
\begin{equation}
  \frac{1}{2} \norm{C^-}^2 (t)  = \frac{1}{2} \norm{C^-}^2 (0) - \int_0^t \left(R^+, C^- \right)(s) \, \diff s \leq 0.
\end{equation}
Since $C^-(\v x, 0) = 0$ and $R^+ C^- \geq 0$ for all $t$, the last inequality holds.
Thus we have that $C^-(\v x, t) = 0$ for all $\v x, t$, and $c = C^+ \geq 0$.
This concludes the proof.

\section{Derivation of manufactured solution}
\label{sec:derivation_manufactured}
Here we derive the analytical solution used to show convergence.
We will assume an incompressible flow where neither density nor permittivity depends on the ion concentrations.

A Taylor--Green vortex flow in the periodic domain $(x, y) \in \Omega = [0, 2\pi] \crossproduct [0, 2\pi]$, is given by
\begin{align}
  \v u &= U(t) (\hat{\v x} \cos x \sin y - \hat{\v y} \sin x \cos y ), \label{eq:manufactured_u} \\
  c_\pm &= c_0 (1 \pm \cos x \cos y \, C(t)). \label{eq:manufactured_c_pm}
\end{align}
Solving the electrostatic problem yields
\begin{align}
  \rho_e &= 2 c_0 \cos x \cos y \, C(t) \\
  V &= \frac{c_0}{\epsilon} \cos x \cos y \, C(t)
\end{align}
which gives a residual of order $O(c_0/\epsilon)$.
We assume the mobilities $K_\pm = D c_\pm$, and the chemical energy function $\alpha(c) = c (\ln c - 1 )$.

The divergence criterion is obtained by taking the divergence of the Navier--Stokes equations with constant density $\rho$:
\begin{equation}
  \rho (\grad \v u)^T : \grad \v u + \div(\rho_e \grad V) = - \laplacian \left( p + \sum_i c_i \right) = - \laplacian p
\end{equation}
Hence, inserting the manufactured solutions \cref{eq:manufactured_u,eq:manufactured_c_pm} yields
\begin{align}
  -\laplacian p &= - \rho U^2(t) (\cos 2x + \cos 2y) - \frac{c_0^2 C^2(t)}{\epsilon} \left( \cos 2x + \cos 2y + 2 \cos 2x \cos 2y \right) \\
                &= -\left( \rho U^2(t) + \frac{c_0^2 C^2(t)}{\epsilon} \right) (\cos 2x + \cos 2y) - \frac{2c_0^2 C^2(t)}{\epsilon} \cos 2x \cos 2y
\end{align}
we find that the pressure is
\begin{equation}
  p = - \frac{1}{4}\left( \rho U^2(t) + \frac{c_0^2 C^2(t)}{\epsilon} \right) (\cos 2x + \cos 2y) - \frac{c_0^2 C^2(t)}{4 \epsilon} \cos 2x \cos 2y
\end{equation}
We have that
\begin{equation}
  \rho_e \grad V = - \frac{c_0^2}{2\epsilon} C^2(t) \left[ \hat{\v x} \sin 2x (1+\cos 2y) + \hat{\v y} (1+\cos 2x) \sin 2y \right]
\end{equation}
and that
\begin{equation}
  \grad p = \frac{1}{2} \left[ \left( \rho U^2(t) + \frac{c_0^2 C^2(t)}{\epsilon} (1 + \cos 2y)\right) \sin 2x \hat{\v x} + \left( \rho U^2(t) + \frac{c_0^2 C^2(t)}{\epsilon} (1 + \cos 2x)\right) \sin 2y \hat{\v y}\right]
\end{equation}
so that
\begin{equation}
  \grad p + \rho_e \grad V = \frac{\rho U^2(t)}{2} \left[ \sin 2x \hat{\v x} + \sin 2y \hat{\v y}\right]
\end{equation}
and since
\begin{align}
  \v u \cdot \grad \v u &= \hat {\v x} (u_x \partial_x u_x + u_y \partial_y u_x) + \hat {\v y} (u_x \partial_x u_y + u_y \partial_y u_y) \\
                        &= -\frac{U^2(t)}{2} \left[\hat {\v x} \sin 2x + \hat {\v y} \sin 2y \right].
\end{align}
Hence, the Navier--Stokes equations give
\begin{align}
  \frac{U'(t)}{U(t)} = - \frac{2 \mu}{\rho}  \implies U(t) = \exp(-2\mu t/\rho ).
\end{align}
Further, the ion transport equations must both be augmented by a carefully chosen source term $q$:
\begin{align}
  \pdt c_\pm + \v u \cdot \grad c_\pm - D \div ( \grad c_\pm + z_\pm c_\pm \grad V) = q(x, y),
\end{align}
where
\begin{equation}
  q (x, y) = \frac{D c_0^2 C^2(t)}{2\epsilon}\left[ \cos 2x + \cos 2y + 2 \cos 2x \cos 2y \right].
\end{equation}
This gives local charge conservation, but a local reaction changes the concentration of both ions.

Insertion gives us that
\begin{equation}
  C(t) = \chi \exp\left(-2D\left(1 + \frac{c_0}{\epsilon}\right) t \right).
\end{equation}
Hence the concentrations decay to the equilibrium concentrations.
Note that $\chi < 1$ in order for the ion concentrations to stay positive.

\subsection*{Acknowledgements}
The authors thank Jonas S.\ Juul for helpful discussions and two anonymous reviewers for comments and suggestions that have helped improve the manuscript.
This project has received funding from the European Union's Horizon 2020 research and innovation program through Marie Curie initial training networks under grant agreement 642976 (NanoHeal), and from the Villum Foundation through the grant ``Earth Patterns.''

\bibliographystyle{elsarticle-num-names}
\bibliography{references}

\end{document}